\documentclass[10pt,conference]{IEEEtran}
\usepackage{cite}
\usepackage{amsmath,amssymb,amsfonts}
\usepackage{algorithmic}
\usepackage{graphicx}
\usepackage{textcomp}
\usepackage{xcolor}
\usepackage[hyphens]{url}
\usepackage{fancyhdr}
\usepackage{hyperref}
\usepackage{multirow}

\usepackage{xspace}
\definecolor{mygreen}{RGB}{0, 128, 0}
\definecolor{myorange}{RGB}{212, 85, 0}

\usepackage[most]{tcolorbox}
\newtcolorbox{commentbox}{
  colback=yellow!10,
  colframe=red!60!black,
  boxrule=0.5pt,
  arc=2pt,
  left=4pt,
  right=4pt,
  top=4pt,
  bottom=4pt,
  fonttitle=\bfseries,
  title=Comment
}
\usepackage{todonotes}

\newcommand{\hpcayear}{}
\begin{document}

%%%%%%%%%%%%%%%%%%%%%%%%%%%%%%%%%%%%%%%%
%%%%%%%%%%%%%% -- UPDATE -- %%%%%%%%%%%%%%%
\newcommand{\hpcasubmissionnumber}{1483}
\title{Flexible In-NAND Cryptographic Processing for Secure Flash Storage} 
%\hpcayear{}}
%%%%%%%%%%%%%%%%%%%%%%%%%%%%%%%%%%%%%%%%

%%%%%%%%%%%%%%%%%%%%%%%%%%%%%%%%%%%%%%%%
%%%%%%%% -- ONLY FOR CAMERA READY -- %%%%%%%%
\def\hpcacameraready{} % Uncomment to build camera-ready version
\newcommand\hpcaauthors{Seock-Hwan Noh$^{1}$,
Hoyeon Lee$^{1}$,
Junkyum Kim$^{2}$,
Junsu Im$^{3}$,
Jay H. Park$^{4}$,\\
Sungjin Lee$^{3}$,
Sam H. Noh$^{5}$,
Yeseong Kim$^{1\dagger}$,
Jaeha Kung$^{6\dagger}$\\
\newline$\dagger$Corresponding authors\\}

\newcommand\hpcaaffiliation{$^{1}$DGIST \quad 
$^{2}$Georgia Tech \quad
$^{3}$POSTECH\\
$^{4}$Samsung Electronics \quad
$^{5}$Virginia Tech \quad
$^{6}$Korea University\\}
\newcommand\hpcaemail{\{nosh3332, lhyzone, yeseongkim\}@dgist.ac.kr,
jun-kyum.kim@gatech.edu,\\
\{junsuim, sungjin.lee\}@postech.ac.kr,
jayhpark530@gmail.com,
samnnoh@vt.edu, jhkung@korea.ac.kr}

%%%%% -- ARTEFACT EVALUATION RESULTS -- %%%%%%
% Uncomment the following based on the badges that were awarded to this paper
%\def\aeopen{}           % The artifact is publically available
%\def\aereviewed{}     % The artefact has been reviewed
%\def\aereproduced{} % The results have been reproduced
%%%%%%%%%%%%%%%%%%%%%%%%%%%%%%%%%%%%%%%%

%%%%%%%%%%%%%%%%%%%%%%%%%%%%%%%%%%%%%
%%%%%%%%%% -- DO NOT MODIFY -- %%%%%%%%%%
%%%%%%%%%%%%%%%%%%%%%%%%%%%%%%%%%%%%%

\author{
  \ifdefined\hpcacameraready
    \IEEEauthorblockN{\hpcaauthors{}}
      \IEEEauthorblockA{
        \hpcaaffiliation{} \\
        \hpcaemail{}
      }
  \else
    \IEEEauthorblockN{\normalsize{HPCA \hpcayear{} Submission
      \textbf{\#\hpcasubmissionnumber{}}} \\
      \IEEEauthorblockA{
        Confidential Draft \\
        Do NOT Distribute!!
      }
    }
  \fi 
}

% Heading and footer for title page
\fancypagestyle{camerareadyfirstpage}{%
  \fancyhead{}
  \renewcommand{\headrulewidth}{0pt}
  \fancyhead[C]{
    \ifdefined\aeopen
    \parbox[][12mm][t]{13.5cm}{\hpcayear{} }    
    \else
      \ifdefined\aereviewed
      \parbox[][12mm][t]{13.5cm}{\hpcayear{} }
      \else
      \ifdefined\aereproduced
      \parbox[][12mm][t]{13.5cm}{\hpcayear{} IEEE International Symposium on High-Performance Computer Architecture (HPCA)}
      \else
      \parbox[][0mm][t]{13.5cm}{\hpcayear{}}
    \fi 
    \fi 
    \fi 
    \ifdefined\aeopen 
      \includegraphics[width=12mm,height=12mm]{ae-badges/open-research-objects.pdf}
    \fi 
    \ifdefined\aereviewed
      \includegraphics[width=12mm,height=12mm]{ae-badges/research-objects-reviewed.pdf}
    \fi 
    \ifdefined\aereproduced
      \includegraphics[width=12mm,height=12mm]{ae-badges/results-reproduced.pdf}
    \fi
  }
  %\fancyfoot[L]{\hpcapubid{} \copyright \hpcayear{} IEEE}
  \fancyfoot[C]{}
}
% Heading and footer for remaining pages
\fancyhead{}
\renewcommand{\headrulewidth}{0pt}
%\fancyhead[C]{\hpcayear{} IEEE International Symposium on
% High-Performance Computer Architecture (HPCA)}

\maketitle

%Enables the camera ready header and footer
\ifdefined\hpcacameraready 
  \thispagestyle{camerareadyfirstpage}
  \pagestyle{empty}
\else
  \thispagestyle{plain}
  \pagestyle{plain}
\fi

\newcommand{\hpcaheight}{0mm}
\ifdefined\eaopen
\renewcommand{\hpcaheight}{12mm}
\fi

%%%%%%%%%%%%%%%%%%%%%%%%%%%%%%%%%%%%%%%%
%%%%%%%% -- PAPER CONTENT STARTS -- %%%%%%%%%

\begin{abstract}
We present FlashVault, an in-NAND self-encryption architecture that embeds a reconfigurable cryptographic engine into the unused silicon area of a state-of-the-art 4D V-NAND structure. FlashVault supports not only block ciphers for data encryption but also public-key and post-quantum algorithms for digital signatures, all within the NAND flash chip.
This design enables each NAND chip to operate as a self-contained enclave without incurring area overhead, while eliminating the need for off-chip encryption.
We implement FlashVault at the register-transfer level (RTL) and perform place-and-route (P\&R) for accurate power/area evaluation.
%achieve physical design closure through placement and routing (P\&R).
Our analysis shows that the power budget determines the number of cryptographic engines per NAND chip.
We integrate this architectural choice into a full-system simulation and evaluate its performance on a wide range of cryptographic algorithms.
%Integrated into a full-system simulation framework, FlashVault is evaluated across a wide range of cryptographic algorithms and data sizes.
Our results show that FlashVault consistently outperforms both CPU-based encryption (1.46$\sim$3.45$\times$) and near-core processing architecture (1.02$\sim$2.01$\times$), demonstrating its effectiveness as a secure SSD architecture that meets diverse cryptographic requirements imposed by regulatory standards and enterprise policies. 
%where cryptographic engines are placed near the SSD controller, 
\end{abstract}

\section{Introduction}
Today, governments and private enterprises leverage personal data to deliver highly customized policies and services~\cite{oecd_data_utilization, mckinsey_data_collect, segment_data_collect}. 
%%In doing so, they often collect sensitive information closely tied to individual privacy, such as national identification numbers, financial records, and biometric data. 
In doing so, they often collect sensitive information closely tied to individual privacy such as financial records and biometric data. 
While such data serves as a critical resource for enabling personalized services, it also becomes a prime target for attacks, with severe consequences if leaked. 
%\samnotes{I feel this paragraph is unnecessarily lengthy. Could be shortened as people as already well aware of the seriousness of breaches.}
% [Modi, feedback] For example, in 2017, U.S. credit bureau Equifax suffered a breach that exposed the social security numbers and credit information of approximately 147 million individuals, resulting in widespread identity theft and financial fraud~\cite{equifax_leak}. 
% [Modi, feedback] Similarly, in 2018, India’s national identification system, Aadhaar, experienced a massive data leak involving billions of biometric and financial records, leading to tangible damages such as online banking fraud and tax refund scams~\cite{aadhaar_leak}. 
For instance, in 2013 and 2017, massive data breaches at two major U.S. corporations, Target (a retailer) and Equifax (a credit reporting agency), respectively, exposed sensitive personal information of millions of individuals~\cite{equifax_leak, target_leak}. 
These incidents resulted in credit card fraud and identity theft, with financial damages reaching hundreds of billions of dollars.
To mitigate such illegal use and data misuse, many countries including the United States, the European Union, Japan, and South Korea have enacted data protection laws that mandate encryption and appropriate technical safeguards when storing personal data on storage devices~\cite{ccpa, gdpr, japan_ppc, korea_act}. 
%user authentication, and access control
In parallel, the industry has adopted standards like TCG Opal~\cite{tcg_ssc} and IEEE 1667~\cite{ieee_std1667} to enhance storage security.

\begin{figure}[t]
    \centering
    \includegraphics[scale=0.70]{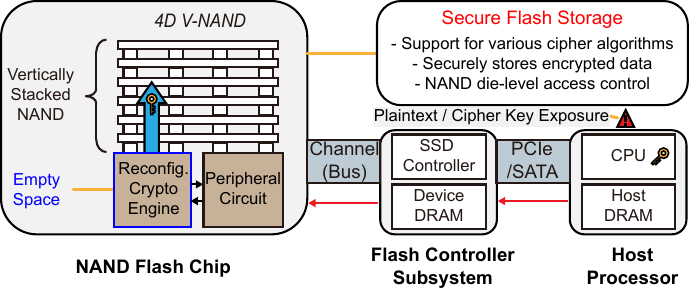}\vspace{-1mm} 
    \caption{ 
    Overview of FlashVault, which embeds a reconfigurable cryptographic engine beneath the 4D V-NAND to perform data encryption and signing directly within NAND. 
    %This in-NAND design shortens the encryption data path and prevents plaintext exposure, unlike host- or controller-level encryption.
    This in-NAND design enables low-latency encryption and prevents plaintext exposure by avoiding off-chip data movement.}
    %\vspace{-2mm}
    \label{fig:flashvault_intro}
   \vspace{-4mm}
\end{figure}

Driven by the increasing adoption of data protection regulations, solid-state drives (SSDs) have adopted the self-encrypting drive (SED) feature to provide hardware-based data protection. 
SEDs encrypt data using dedicated hardware engines embedded in the device and offer data integrity and access control through digital signatures and authentication mechanisms. 
However, due to the following limitations, they fall short of addressing evolving security requirements.\newline
\textbf{[Limitation 1]} \textit{Modern SSDs support only a limited set of cryptographic algorithms.}
In accordance with data protection laws, many countries require or recommend that public institutions and private enterprises handling personal information adopt national standard cryptographic algorithms~\cite{ccpa, gdpr, japan_ppc, korea_act}.
However, current commercial SSDs currently support only specific algorithms, e.g., AES, SHA, and RSA~\cite{samsung_secure_firmware, micron_secure_firmware, ibm_tls, SK_Hynix_ssd_product}.
% [Modi, feedback] As a result, when unsupported algorithms are required, encryption needs to be offloaded to the software stack on the host CPU~\cite{snia_cpu}.
As a result, whenever unsupported algorithms are required, encryption is offloaded to the software stack on the host CPU~\cite{snia_cpu}.
%\samnotes{How often does this kind of situation arise?}
This software-based encryption incurs I/O latency and may increase system-wide response time~\cite{d_shield}.
Such delays not only degrade the user experience in public and enterprise services but also cause significant disruption in latency-sensitive applications, such as high-frequency trading systems~\cite{hft_damage} and decentralized exchanges~\cite{blockchain_damage}. \newline\textbf{[Limitation 2]} \textit{Unencrypted data may be stored in NAND flash memory.} Modern SSDs are equipped with data encryption and SED features~\cite{firmware_attack, tcg_sed}. 
However, the activation of these features is typically determined by user configuration~\cite{dell_sed, qnap_sed}.
Thus, if encryption is not enabled, unencrypted data (i.e., plaintext) is stored in NAND flash memory.
Even when encryption is enabled, residual traces of cryptographic keys may remain in NAND flash due to garbage collection (GC) and wear-leveling (WL) during the key revocation process~\cite{wear_leveling_attack1, adaptive_privacy_ssd}.
In addition, if a power loss occurs, plaintext data residing in the DRAM of the SSD may be flushed to NAND flash~\cite{power_loss_date, power_loss_fast}.
\newline\textbf{[Limitation 3]} \textit{NAND flash lacks architectural protections against software-based and physical attacks.} Since commercial SSDs do not support trusted execution environments within their NAND flash chips, attackers can manipulate the flash translation layer (FTL) mapping tables to distort the data organization or analyze FTL logs to retrieve or tamper with residual data that have been logically deleted but remain physically present~\cite{iceclave}.
In addition, attackers can connect NAND readers to the flash chip to extract unencrypted data, or remnants of data left behind due to GC and WL~\cite{firmware_attack}.
Beyond these device-level threats, when encryption operations are offloaded to the host CPU, the risk of exposure to attacks increases further, weakening overall system security.
For instance, plaintext data transmitted between the SSD and host may be intercepted through bus probing~\cite{anti_bus_probing, anti_bus_probing2}, and cryptographic keys may be inferred or stolen through power analysis~\cite{power_attack, power_attack2} or electromagnetic analysis~\cite{em_attack, em_attack2} during CPU-based encryption processes.

\begin{figure}[t]
    \centering
    \includegraphics[scale=0.72]{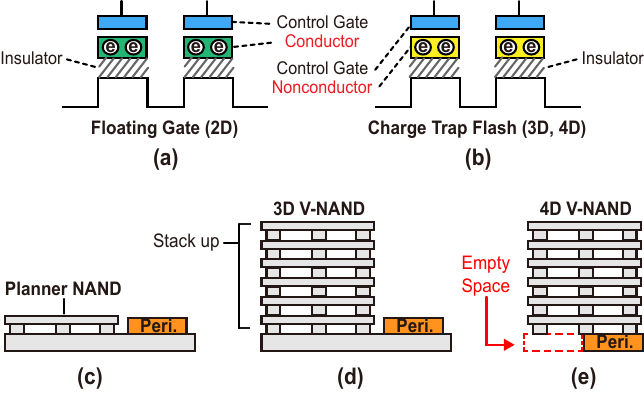}\vspace{-1mm} 
    \caption{Device structures of (a) a floating gate transistor and (b) a charge trap flash. Structures of (c) 2D, (d) 3D and (e) 4D NAND flash memories.}  
    \label{fig:3d_4d_ssd}
    \vspace{-4mm}
\end{figure}

To address these limitations of existing SSDs, we propose \textit{\underline{FlashVault}}, the first in-NAND self-encryption architecture that supports a wide range of cryptographic algorithms directly within NAND flash (Fig.~\ref{fig:flashvault_intro}). 
%\samnotes{Isn't FlashVault limited to 4D V-NAND, which is next generation NAND? If it is, it should be clearly conveyed here before discussing the features.}
It is specifically designed for modern 4D V-NAND architectures, which adopt a peripheral-under-cell structure with unused space beneath the memory array.
By embedding reconfigurable cryptographic engines into this unused space, FlashVault enables a fast and secure in-NAND enclave without incurring additional area overhead.

The specific architectural features of FlashVault are: 
%as follows:
\begin{enumerate}\vspace{-1mm}
    \item \textbf{Reconfigurable cryptographic engine:} FlashVault integrates a reconfigurable cryptographic engine that supports a variety of block ciphers and public-key cryptographic algorithms (PKC). It also supports post-quantum cryptography (PQC), providing the architectural flexibility needed to address emerging security threats posed by quantum computing.
    \item \textbf{Utilization of the unused area beneath 4D V-NAND:} In FlashVault, the cryptographic engine is placed in the unused silicon area beneath the 4D V-NAND array, enabling secure functionality without incurring additional area overhead. By performing encryption and decryption near the memory array, FlashVault ensures data confidentiality and integrity at the NAND chip level. Furthermore, by internalizing operations that were traditionally offloaded to the CPU, FlashVault eliminates potential attack surfaces along the data path to the host.
    \item \textbf{Die-level shared integration architecture:} %FlashVault enables high-throughput encryption and decryption by integrating cryptographic engines in parallel at the NAND plane level.
    FlashVault supports high-throughput encryption and decryption by employing a pipelined, die-level shared integration of cryptographic engines within each NAND chip.
    This architecture facilitates the deployment of computationally intensive cryptographic algorithms that were previously impractical for real-time secure applications.
\end{enumerate}

\vspace{-1mm}
%This paper focuses on the architecture of FlashVault and its performance and energy efficiency.
This paper focuses on the architecture of FlashVault and its performance.
%While security functionality is a fundamental requirement driven by legal and industrial standards, performance and energy efficiency are also critical factors for the practical deployment of secure storage systems in real-world environments~\cite{storage_energy_latency, iceclave, rif}.
While security functionality is a fundamental requirement driven by legal and industrial standards, performance is also a critical factor for the practical deployment of secure storage systems in real-world environments~\cite{storage_energy_latency, iceclave, rif}.
%While security functionality is a fundamental requirement driven by legal and industrial standards, performance is also critical factors for the practical deployment of secure storage systems in real-world environments~\cite{iceclave, rif}.
Moreover, the structural characteristics of FlashVault inherently mitigate security vulnerabilities observed in conventional systems (Section~\ref{sec:threat_model}).

%The remainder of this paper is organized as follows: Section~\ref{sec:background} provides the background for understanding this work.
%Section~\ref{sec:motivation} discusses the need for supporting diverse cryptographic algorithms and the limitations of CPU-based encryption.
%Sections~\ref{sec:design_requirement} and~\ref{sec:overall_architecture} describe the design considerations and overall architecture of FlashVault.
%Section~\ref{sec:Microarchtiecture} presents the microarchitecture of the integrated cryptographic engine.
%Section~\ref{sec:evaluation} evaluates the architectural benefits of FlashVault.
%Finally, Section~\ref{sec:conclusion} concludes the paper.

\begin{figure}[t]
    \centering
    \includegraphics[scale=0.77]{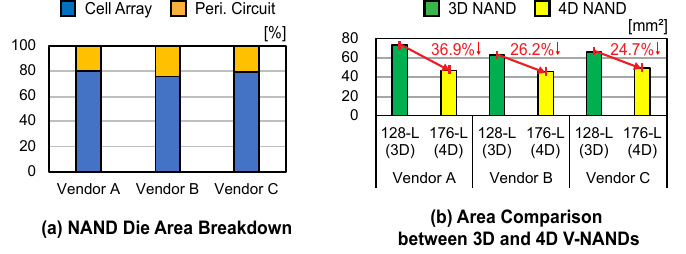}\vspace{-1mm} 
    \caption{(a) Area breakdown of a 512 Gb 3D V-NAND chip into the cell array and peripheral circuits for three major vendors. 
    Peripheral circuits include components such as row decoders, page buffers, and charge pumps. 
    (b) Comparison of die sizes between 128-layer 3D NAND and 176-layer 4D NAND flash from the same vendors. 
    Sources:~\cite{pif, tech_insight_survey, hynix_die_area, Hynix_4d_nand, 3d_4d_comparison_nand}.}  
    \label{fig:3d_4d_area_comparison}\vspace{-4mm}
\end{figure}

%\vspace{-1mm}
\section{Preliminaries}\label{sec:background}
% \vspace{-1mm}
\subsection{NAND Flash Memory Technology}\label{sec:4d_nand}%\vspace{-1mm}
The memory cells in early-generation SSDs are based on floating gate transistors (FGTs). 
An FGT is a complementary metal-oxide-semiconductor (CMOS) technology capable of holding electrical charges in an isolated silicon conductive layer (Fig.~\ref{fig:3d_4d_ssd}-(a))~\cite{overview_nvm_tech}. 
Early generation flash memory arrays have a 2D planar structure with FGTs placed side-by-side on the die (Fig.~\ref{fig:3d_4d_ssd}-(c)). 
However, as the process node has scaled down, FGTs encountered fabrication challenges, i.e., charge leakage caused by the thin conductive layer and intensified cell-to-cell interference~\cite{cell_interference, modern_ssd_survey}. 
To alleviate this limitation, FGTs were gradually superseded by transistors utilizing charge trap technology. 
These newer transistors employ a non-conductive silicon layer as a control gate, highly effective at retaining electrical charges (Fig.~\ref{fig:3d_4d_ssd}-(b))~\cite{overview_nvm_tech, modern_ssd_survey}. 
%\samnotes{In Fig. 2(b): tranp $$->$$ trap}
Moreover, to meet demands for higher density, memory suppliers developed a 3D NAND structure (Fig.~\ref{fig:3d_4d_ssd}-(d)), featuring vertically stacked charge trap transistors alongside the peripheral circuit on the same integrated chip~\cite{samsung_3d_vnand, micron_3d_nand, Hynix_3d_vnand}. 
This 3D architecture significantly increased storage capacity by adding more layers achieving lower cost/bit.
More recently, the industry has introduced 4D NAND flash products\footnote{In the industry, 4D V-NAND is also referred to as PUC (Periphery Under Cell) by SK hynix, CUA (CMOS Under Array) by Micron, and COP (Cell Over Periphery) by Samsung~\cite{tech_insight_survey}.}, which adopt an advanced vertical integration scheme to improve bit density (Fig.~\ref{fig:3d_4d_ssd}-(e)).
These offer a reduced chip area compared to 3D NAND flash by positioning the peripheral circuit underneath the memory array~\cite{peri_area}.
Fig.~\ref{fig:3d_4d_area_comparison}-(a) illustrates the area composition of 512 Gb-class 3D V-NAND chips from three major memory manufacturers, indicating that peripheral circuits account for approximately 20$\sim$24\% of the total chip area.
%\samnotes{Fig 3: vender $->$ vendor}
By placing the peripheral circuits under the memory array, the 4D structure not only reduces the die footprint by a comparable amount (Fig.~\ref{fig:3d_4d_area_comparison}-(b)), but also leaves a portion of the bottom silicon layer unoccupied, since the relocated circuits are smaller in area than the memory array above.
Motivated by this architectural characteristic, there have been efforts to leverage the resulting unused area by integrating auxiliary logic circuits, such as pattern matchers~\cite{pif}, data processing units~\cite{4d_vnand_data_processing}, and neuron-inspired circuits~\cite{4d_vnand_neuron_circuit}.
%\noh{This work proposes to use this area for ...}\samnotes{How about something like this here?}
This work leverages this area to embed reconfigurable cryptographic engines, enabling fast and secure near-data cryptographic processing.

\begin{figure}[t]
    \centering
    \includegraphics[scale=0.62]{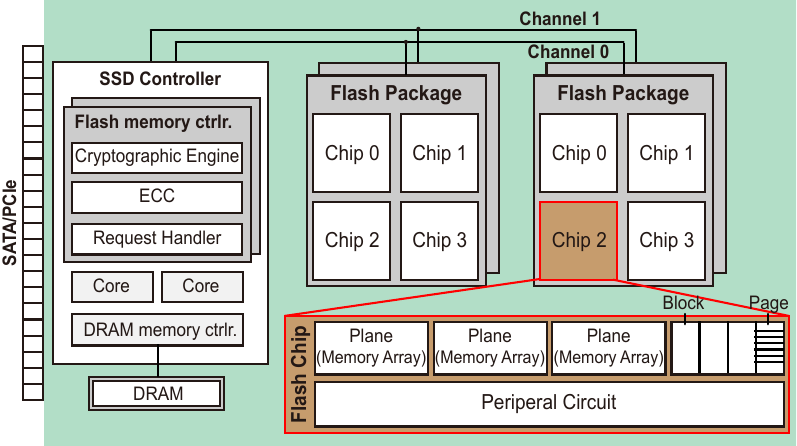}%\vspace{-1mm} 
    \caption{Internal architecture of modern SSDs.}    
    \label{fig:SSD_overall_architecture}
    \vspace{-4mm}
\end{figure}

%\vspace{-1mm}
\subsection{Internal Architecture of SSDs}\label{sec:SSD_internal_architecture}
%\vspace{-1mm}
Fig.~\ref{fig:SSD_overall_architecture} illustrates the typical architecture of modern SSDs~\cite{modern_ssd_isca, modern_ssd_survey, modern_ssd_amber, decoupled_ssd}.
An SSD consists of NAND flash packages, an SSD controller, and DRAM. 
Each NAND flash package contains multiple flash chips, each housing several memory planes. 
These planes are further divided into blocks and pages. 
A block is the smallest erasable unit, typically holding 1,024 pages. 
A page, the basic unit for read and program/write operations, usually has a size between 4 KB and 16 KB. 
Data read from or programmed to the flash memory is managed by a flash memory controller. 
This controller orchestrates data distribution to prevent bottlenecks among NAND packages sharing a byte-wide channel. 
%It also corrects errors in the data read from the memory array and handles encryption and decryption processes of the AES cipher through an AES cryptographic engine.
It also performs error correction on data read from the memory array and performs encryption and decryption of specific cipher algorithms, i.e., AES, SHA, and RSA, via a dedicated cryptographic engine.
The SSD controller, comprising 1 to 4 flash memory controllers, is equipped with an embedded multicore processor. 
%\samnotes{'a couple of ...controller'? Just two? not many?}
This processor operates the internal firmware known as the flash translation layer (FTL). 
The FTL is responsible for translating addresses between the logical memory space and the physical flash page, performing garbage collection (GC) to reclaim invalid or empty spaces, and executing wear leveling (WL) to ensure even memory usage across the memory blocks. 
The DRAM is utilized for storing a virtual-to-physical address mapping table and the firmware code of the FTL.

%\vspace{-1mm}
\subsection{Threat Model}\label{sec:threat_model}
%\vspace{-1mm}
Modern commercial SSDs provide hardware-based data protection through SED feature, typically employing AES encryption with a dedicated engine in the SSD controller~\cite{firmware_attack, tcg_sed}.
These controllers also support cryptographic primitives such as SHA and RSA for integrity verification and firmware authentication~\cite{samsung_secure_firmware, micron_secure_firmware, ibm_tls, SK_Hynix_ssd_product}. 
In current SSD architectures, encryption is typically performed in one of two ways: (i) by a hardware engine integrated into the SSD controller or
(ii) in software on the host CPU when the required algorithm is not supported by the controller.

%When encryption is handled by the SSD controller, the NAND flash chip stores the encrypted data received from the controller~\cite{iceclave}.
When the \textit{SSD controller} performs encryption, the NAND flash stores encrypted data sent from the controller~\cite{iceclave}.
%However, the presence of encrypted data in NAND does not fully eliminate potential security risks.
However, storing encrypted data in NAND does not eliminate all security risks.
%For instance, during key revocation process, residual key data may remain in NAND flash due to WL or GC~\cite{wear_leveling_attack1, adaptive_privacy_ssd}.
%\samnotes{The lower half of this paragraph is repetitive. Most have already been discussed in the introduction. You might just want to refer to the limitations mentioned earlier.}
% [modi, feedback] For example, during key revocation, leftover key data may remain in NAND due to WL and GC~\cite{wear_leveling_attack1, adaptive_privacy_ssd}.
%In addition, under power loss conditions, plaintext data residing in the DRAM of the SSD may be flushed to the NAND flash~\cite{power_loss_date, power_loss_fast}.
% [modi, feedback] If power is lost, plaintext data stored in the DRAM of the SSD may be flushed to NAND~\cite{power_loss_date, power_loss_fast}.
%These residual keys or plaintext fragments can be extracted through physical attacks using equipment such as flash readers~\cite{firmware_attack}.
%[ modi, feedback]These residual keys or plaintext fragments can be extracted via physical attacks using flash readers~\cite{firmware_attack}.
For example, remnants of encryption keys or plaintext data may persist in NAND flash due to WL, GC~\cite{wear_leveling_attack1, adaptive_privacy_ssd}, or unexpected power loss~\cite{power_loss_date, power_loss_fast}. 
These residual traces may be extracted through physical attacks using flash readers~\cite{firmware_attack}.
% [modi, feedback] Moreover, off-the-shelf NAND chips lack built-in authentication and access control mechanisms, leaving them vulnerable to unauthorized access through malicious firmware.
%Commercial SSDs remain vulnerable to firmware-based attacks, where adversaries may inject modified firmware to tamper with or exfiltrate stored data~\cite{firmware_attack}.
% [modi, feedback]This allows adversaries to inject modified firmware into SSDs to tamper with or exfiltrate data~\cite{firmware_attack}.
%stored data~\cite{firmware_attack}.
Moreover, off-the-shelf NAND chips lack built-in access control mechanisms, leaving them vulnerable to unauthorized access through malicious firmware~\cite{firmware_attack}.

When encryption is performed by the \textit{CPU}, sensitive data can be exposed to adversaries through probing attacks on the data bus~\cite{anti_bus_probing, anti_bus_probing2}. 
Attackers may also analyze power consumption patterns~\cite{power_attack, power_attack2} or electromagnetic emissions~\cite{em_attack, em_attack2} using sensors or oscilloscopes to infer secret keys or other sensitive information. 
Furthermore, software-based encryption is at risk of microarchitectural attacks, such as cache timing side channels, which can leak sensitive intermediate results that depend on cryptographic secrets~\cite{cache_attack1, cache_attack2}.
%which can leak secret-dependent data during cryptographic operations~\cite{cache_attack1, cache_attack2}.}

FlashVault mitigates these threat models by placing dedicated cryptographic engines beneath the memory array of 4D V-NAND.
These engines are physically located in the deep silicon layer underneath the stacked array, effectively removing externally accessible surfaces and preventing physical access-based attacks.
FlashVault ensures that only encrypted data remains in NAND even under conditions such as WL, GC, or power loss through in-NAND encryption.
When used in conjunction with the encryption built into the SSD controller, FlashVault creates a secure enclave that spans the entire data path from the controller to the NAND flash media.
%When used in conjunction with the built-in encryption mode of the SSD, FlashVault enables double encryption,
%\samnotes{What do you mean by 'double encryption'?}
%ensuring that data remains protected even during WL or GB processes.
Furthermore, FlashVault supports PKC and PQC directly within the NAND layer, providing on-die authentication to block unauthorized firmware accesses.
%and reduce potential threats.
Finally, by eliminating the need to offload cryptographic workloads to the host CPU, FlashVault inherently avoids a wide range of CPU-side vulnerabilities, including software attacks and side-channel leakage.

%\vspace{-1mm}
\section{Motivation}\label{sec:motivation}
% \vspace{-1mm}
%\samnotes{I feel there is a lot of redundancy here with the introduction and the background. Could/should have been much more compact.}
\subsection{Support for Diverse Cryptographic Algorithms}\label{sec:moti_diverse_algorithms}
%\textbf{1) Block Cipher Algorithm: }

\subsubsection{Block Cipher Algorithm}\label{sec:moti_block_cipher}
Modern SSDs provide a strong security level by employing block cipher algorithms that enable fast encryption and decryption to protect data.
%Most commercial self-encrypting SSDs support fixed block cipher algorithms such as AES~\cite{ieee_aes_std, intel_aes, kingston_standard}, in accordance with international security standards for storage devices, including FIPS 140-2/3~\cite{fips_140_2, fips_140_3}, TCG Opal~\cite{tcg_ssc}, and IEEE 1619~\cite{ieee_aes_std}.
Most commercial self-encrypting SSDs support fixed block cipher algorithms such as AES~\cite{ieee_aes_std, intel_aes, kingston_standard}, following storage security standards such as FIPS 140-2/3~\cite{fips_140_2, fips_140_3}, TCG Opal~\cite{tcg_ssc}, and IEEE 1619~\cite{ieee_aes_std}.
However, modern security requirements are becoming increasingly diverse.
%In particular, to protect sensitive national information and ensure cryptographic sovereignty in line with national security demands, many advanced countries have enacted and strengthened data protection laws. 
To protect data vital to national security and public welfare, and maintain cryptographic sovereignty, many countries have enacted stricter data protection laws.
Accordingly, they mandate or recommend the use of national standard block cipher algorithms to protect sensitive data (e.g., personal identifiers, financial records, and military information) stored in storage media.
For example, the United States, the European Union, South Korea, China and Russia require the use of AES/Triple DES~\cite{cmvp}, AES~\cite{gdpr}, ARIA/SEED/HIGHT~\cite{kcmvp_example}, SM4~\cite{sm4_report}, and GOST~\cite{gost}, respectively.
Japan and Turkey recommend Camellia~\cite{japan_camellia} and KET~\cite{kvkk} for securing public sector data.
% [Revision] Industry sectors have also begun to adopt various block ciphers depending on their distinct security and compliance needs.
%  [Revision] For example, sectors such as finance and healthcare, where high security and regulatory compliance are critical, typically prioritize the use of national standard algorithms~\cite{HIPAA_rule, pci_financial}.
%Meanwhile, global service providers that operate across multiple jurisdictions increasingly support a broader range of algorithms, such as SM4, 3DES, and Twofish, alongside AES to meet region-specific legal and technical requirements~\cite{hauwei_encryption, ndvpn2}.
This growing diversity of security demands indicates that relying on a single fixed block cipher algorithm in SSDs is inadequate to address today’s heterogeneous requirements.

%\noindent\textbf{2) Public-Key Cryptographic Algorithms:} 
%\vspace{-1mm}
\subsubsection{Public-Key Cryptographic Algorithm}\label{sec:moti_puk}
Recent storage devices are equipped with advanced security features based on data integrity and digital signatures.
%to meet the evolving security demands of computing environments. 
Storage security standards such as TCG Opal~\cite{tcg_ssc} and NVMe-oF specify public-key-based authentication mechanisms as core components, leading modern SSDs to support a limited set of public-key cryptographic algorithms such as RSA and NIST ECC-crypto\footnote{ECC-crypto refers to Elliptic Curve Cryptography.} (e.g., secp256r1)~\cite{samsung_secure_firmware, micron_secure_firmware, ibm_tls, SK_Hynix_ssd_product}.
However, these fixed-function designs inherently lack flexibility to accommodate diverse regional security requirements.
For instance, Germany’s Federal Office for Information Security recommends ECC-crypto based on Brainpool curves~\cite{ecc_bsi}, while China promotes its domestic SM2 algorithm in industry~\cite{china_sm2}.
However, commercial SSDs generally do not support such region-specific cryptographic algorithms, limiting their applicability in regions with specific regulatory requirements.

Moreover, as storage devices are entrusted with preserving sensitive data over long periods, ensuring their long-term security against cryptographic threats is critical.
From this perspective, the advent of quantum computing poses a threat to existing public-key cryptosystems, since quantum computers can solve the underlying mathematical problems such as integer factorization and discrete logarithms using Shor’s algorithm~\cite{quantum_threat}.
This raises the concern that today’s widely adopted public-key security mechanisms may become ineffective in the near future. 
%To address the looming threat, NIST has been leading the standardization efforts for post-quantum PKC over the past several years.
%As a result of this process, Dilithium, FALCON and SPHINCS+ have been officially selected as the standard algorithms for digital signatures~\cite{nist_pqc_altorithms}.
To address this, NIST has led post-quantum standardization and selected Dilithium, FALCON, and SPHINCS+ as digital signature standards~\cite{nist_pqc_altorithms}. 
In response, supporting these PQC algorithms is emerging as a critical requirement for future storage systems that can withstand future cryptographic threats.
%that must remain resilient in evolving security landscapes.

%\vspace{-1mm}
\subsection{Limits of CPU-based Encryption}\label{sec:moti_nand_sed}
% \vspace{-1mm}
Modern SSDs have evolved beyond simple data storage devices into intelligent security platforms equipped with various built-in protection mechanisms. 
In addition to encrypting data to ensure confidentiality, these devices also support integrity-focused features such as secure boot and tamper-proof logging, which perform signature verification and detect unauthorized modifications during system initialization and log management.
To implement these security functions effectively, different cryptographic algorithms may be selected depending on the application domain and security priorities.
For example, SSDs deployed in security-critical sectors, such as defense, may employ robust signature schemes such as ECDSA~\cite{ecdsa} or upcoming post-quantum algorithms to ensure high assurance.
On the other hand, SSDs in latency-critical environments like financial trading systems may use lightweight block ciphers such as HIGHT~\cite{HIGHT} or Camellia~\cite{camellia} to minimize encryption latency while still ensuring data protection.

However, off-the-shelf SSDs typically support only a limited set of algorithms at the hardware level~\cite{samsung_secure_firmware, micron_secure_firmware, ibm_tls, SK_Hynix_ssd_product}, requiring other cryptographic operations to be offloaded to the host processor~\cite{snia_cpu}.
This CPU-based processing often incurs significant latency, making it difficult to meet the stringent timing requirements of practical applications.
% [revision] As a result, while a broader range of algorithms may be considered in theory, real-world deployment is often constrained by such latency overhead.
To mitigate this issue, it is necessary to minimize reliance on host-side processing in order to reduce the latency overhead caused by offloading cryptographic operations to the host processor.
%FlashVault is designed to meet this demand through In-NAND self-encryption that enables energy-efficient and high-performance execution of various cryptographic algorithms.
FlashVault addresses this challenge by supporting in-NAND self-encryption, enabling high-performance execution of diverse cryptographic algorithms.
This section motivates the need for in-NAND processing by analyzing the inefficiencies of CPU-based encryption in real-world scenarios.

%\vspace{-1mm}
\subsubsection{Experimental Setup}\label{sec:moti_exp_setup}
% [Revision] To quantitatively evaluate the performance overhead of CPU-based encryption, we measured the latency and power consumption of various cryptographic algorithms executed on a real hardware platform. 
% [Revision] The experiments were conducted on a modern desktop system equipped with an Intel Core i7-13700K processor (16 cores), 32 GB of DDR5-5600 memory, and a 2TB SK Hynix Gold P31 M.2 NVMe SSD (PCIe 3.0 $\times$ 4), running Ubuntu 22.04 LTS.
To quantitatively evaluate the performance overhead of CPU-based encryption, we measured the latency of various cryptographic algorithms on a desktop system with an Intel Core i7-13700K (16 cores), 32 GB DDR5-5600 memory, and a 2TB SK Hynix Gold P31 M.2 NVMe SSD (PCIe 3.0 ×4), running Ubuntu 22.04 LTS.
Block ciphers and public-key cryptographic algorithms were implemented using the Botan library~\cite{botan}, and PQC algorithms were evaluated using the openssl speed benchmark tool in conjunction with liboqs~\cite{liboqs} and OpenSSL~\cite{openssl}. 
For block ciphers, we used fixed 4 KB input messages, whereas public-key and PQC algorithms were evaluated across a range of input sizes.
%In particular, the experiments on public-key algorithms were based on scenarios modeling the verification process of secure boot and the signature validation in security log systems such as Tamper-Proof Logging.
In particular, the experiments on public-key algorithms were based on scenarios modeling the verification process of secure boot and the signature validation used in security logging systems such as tamper-proof logging.
As part of the analysis of secure boot latency, we used the latency value reported in~\cite{texas_boot_size} for the boot code loading process.
%Latency and power consumption were measured using the \texttt{clock\_gettime()} function and the Intel \texttt{RAPL interface}, respectively. 
Latency was measured using the \texttt{clock\_gettime()} function, with \texttt{fsync()} and \texttt{posix\_fadvise()} used to reduce the effects of OS-level file caching.
%% -- [Evaluation part] Power consumption was collected via Intel \texttt{RAPL interface} for the CPU and DRAM, while SSD power consumption, which cannot be measured by RAPL, was estimated using Micron’s SSD NAND flash Power Calculator~\cite{power_caculator}.
Each algorithm was evaluated 1,000 times per input size, and we report the average latency. 
%and energy consumption.
%To assess the real-world implications of energy consumption, the measured energy values were converted into operational electricity cost and carbon emissions. 
% [Ealuation part] To assess the real-world implications of energy consumption, the annual energy usage of a 1,000-CPU system was extrapolated and converted into estimated electricity cost and carbon emissions.
% [Evaluation part] The electricity cost was estimated based on the average industrial electricity rate in the United States (approximately \$0.12/kWh)~\cite{power_cost}, and the carbon emissions were calculated using the carbon intensity factor published by the International Energy Agency (IEA), approximately 0.385 kgCO$_{2}$/kWh~\cite{co2_emission}.

\begin{figure}[t]
    \centering
    \includegraphics[scale=0.96]{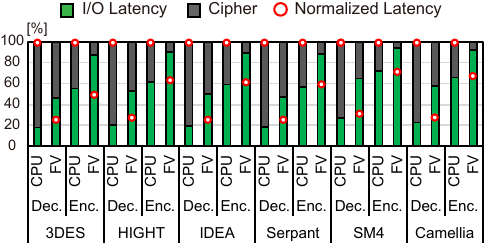}\vspace{-1mm} 
    \caption{Normalized latency breakdown of decryption (Dec.) and encryption (Enc.) using six representative block cipher algorithms on CPU and FlashVault (FV). 
    Each bar shows I/O latency (green), cipher latency (gray), with red circles indicating total latency normalized to the CPU implementation.}    
    \label{fig:moti_bloci_cipher}
    \vspace{-1mm}
\end{figure}

\begin{figure}[t]
    \centering
    \includegraphics[scale=0.98]{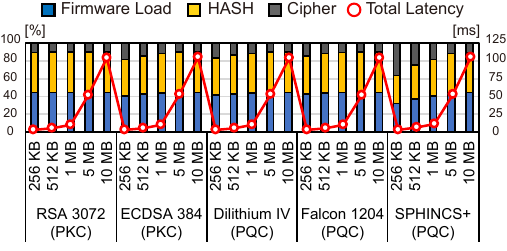}\vspace{-2mm}  
    \caption{Latency breakdown of public-key cryptographic verification operations during secure boot, evaluated with various input sizes. 
    The stacked bars show relative contributions of boot image loading (blue), hash computation (orange), and PKC/PQC operations (gray), with red circles indicating the total latency.}    
    \label{fig:moti_puk_verify}
    \vspace{-3mm}
\end{figure}

%\vspace{-1mm}
\subsubsection{Performance of Block Cipher Algorithms}\label{sec:moti_performance_block_cipher}
% \vspace{-1mm}
%\samnotes{I would have liked to see the absolute latency numbers for the CPU case. Also, showing FlashVault numbers without even having described anything about it looks a bit awkward.}
Fig.~\ref{fig:moti_bloci_cipher} shows the latency breakdown for six representative block ciphers (3DES~\cite{3des}, HIGHT~\cite{HIGHT}, IDEA~\cite{idea}, Serpent~\cite{serpant}, SM4~\cite{sm4_report}, Camellia~\cite{camellia}) executed in CTR mode, measured under both CPU and FlashVault.
Although FlashVault embeds multiple cryptographic engines within each NAND flash chip, Fig.~\ref{fig:moti_bloci_cipher} reports the result based on a single-block-per-chip configuration to focus on the impact of where encryption occurs.
%\footnote{\revision{While FlashVault is capable of integrating multiple cryptographic blocks per NAND flash chip, this figure presents results based on a single-block-per-chip configuration to examine the effect of encryption placement.}} (the experimental setup for FlashVault is described in Section~\ref{sec:archi_evaluation_methodology}.)
The experimental setup for FlashVault is described in Section~\ref{sec:archi_evaluation_method}.
Each bar in the figure is divided into encryption latency and I/O latency to illustrate their respective contributions to the total processing time.
%In the CPU-based implementation, the serial execution of encryption operations and DRAM accesses during computation account for a significant portion of the overall latency.
In the CPU-based implementation, limited concurrency in cryptographic processing and the frequent data transfers between the CPU and host DRAM contribute significantly to the overall latency.
In contrast, when cryptographic operations are performed inside in-NAND flash, as implemented in FlashVault, this latency can be mitigated by eliminating the host-side cryptographic processing.
Consequently, FlashVault achieves an average latency reduction of 64.2\% for decryption and 37.0\% for encryption compared to CPU-based processing. 
% This demonstrates that in-NAND self-encryption can improve in-line encryption performance and enable efficient security processing at the storage level.

\subsubsection{Latency of PKC and PQC Verification} %Fig.~\ref{fig:moti_puk_verify} shows the latency breakdown of the signature verification process using well-known PKC and PQC algorithms during secure boot, evaluated with boot image sizes ranging from 256 KB to 10 MB.
Fig.~\ref{fig:moti_puk_verify} shows the signature verification latency using well-known PKC and PQC during secure boot for boot images from 256 KB to 10 MB.
The verification sequence starts by loading the firmware image from boot ROM or NAND flash, computes a cryptographic hash, e.g., SHA~\cite{sha}, and verifies the resulting digest using a digital signature algorithm, e.g., PKC or PQC.
%The verification sequence begins by loading the firmware image from boot ROM or NAND flash, followed by computing a cryptographic hash (e.g., SHA~\cite{sha}), and then verifying the resulting digest using a digital signature algorithm (e.g., a PKC or PQC algorithm)~\cite{secure_boot_qualcomm, secure_boot_nand}.
% MAsk ROM, eFuse
The input to the signature verification is relatively small, as it is a fixed-length digest produced by the hash function. 
%However, since the boot image itself ranges from several hundred KB to tens of MB, image loading and hash computation account for a large portion of the total verification latency.
However, since boot images range from hundreds of KB to tens of MB, image loading and hash computation account for a large portion of the total verification latency.
Meanwhile, the overall verification latency varies with the size of the boot image, increasing as the image size grows.
As shown in the figure, when a 10 MB image is used, the signature verification time exceeds 100 ms across all cryptographic algorithms~\cite{secure_boot_100ms}.
This result indicates that it is difficult to meet the constraint imposed by commercial hardware implementations, which require boot-time verification to be completed within 100 ms after system power stabilization (i.e., \texttt{power\_good}).
%\footnote{This constraint has led modern SSD products to support a PCK algorithm, most commonly RSA, through the encryption engine in the SSD controller~\cite{samsung_secure_firmware, micron_secure_firmware, ibm_tls, SK_Hynix_ssd_product}.}
Moreover, as modern high-end SSDs often adopt firmware images larger than 10 MB~\cite{samsung_firmware_size}, meeting the timing constraint using CPU-based verification becomes even more challenging.

\begin{figure}[t]
    \centering
    \includegraphics[scale=0.98]{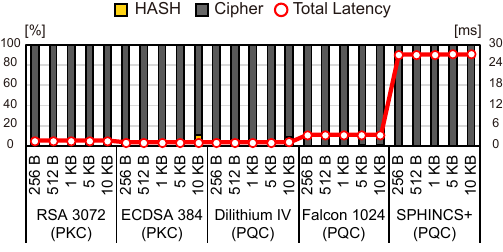}\vspace{-2mm}
    \caption{Latency breakdown of public-key cryptographic signing operations in tamper-proof logging evaluated across various input sizes.
    Each stacked bar shows the contributions of hash computation (orange) and PKC/PQC operations (gray), and red circles indicate the total latency.}    
    \label{fig:moti_puk_sign}
    \vspace{-3mm}
\end{figure}

\begin{table*}[t]
\centering
%\begin{threeparttable}
\caption{Primitive operations of representative cryptographic algorithms}\vspace{-2mm}
\label{tab:cipher_operation_summary}
\scalebox{0.8}{
\begin{tabular}{c|c|c|c|c|c|clllllll}
\hline\hline
\textbf{Category}                                                                & \textbf{Algorithm} & \textbf{\# of Rounds$^*$} & \textbf{\begin{tabular}[c]{@{}c@{}}Key Length \\ {[}bit{]}\end{tabular}} & \textbf{\begin{tabular}[c]{@{}c@{}}Input Unit Size$^\dag$ \\ {[}bit{]}\end{tabular}} & \textbf{\begin{tabular}[c]{@{}c@{}}Processing \\ Granularity$^\ddagger$ {[}bit{]}\end{tabular}} & \textbf{Required Primitive Operations}                                                                 \\ \hline\hline
\multirow{7}{*}{\textbf{\begin{tabular}[c]{@{}c@{}}Block\\ Cipher\end{tabular}}} & AES~\cite{aes_first}                & 10 / 12 / 14            & 128 / 192 / 256                                                          & 128                                                               & 8                                                                                    & XOR, S-Box, Shift, Multiplication                                                                      \\ \cline{2-7} 
                                                                                 & 3DES~\cite{3des}               & 48                    & 112 / 168                                                                & 64                                                                & 4                                                                                    & XOR, S-Box, Permutation                                                                                \\ \cline{2-7} 
                                                                                 & IDEA~\cite{idea}               & 8.5                   & 128                                                                      & 64                                                                & 16                                                                                   & XOR, ModAdd, ModMult                                                                                   \\ \cline{2-7} 
                                                                                 & Serpant~\cite{serpant}            & 32                    & 128 / 192 / 256                                                          & 128                                                               & 4                                                                                    & XOR, S-Box, Shift, Permutation                                                                         \\ \cline{2-7} 
                                                                                 & HIGHT~\cite{HIGHT}              & 32                    & 128                                                                      & 64                                                                & 8                                                                                    & XOR, Shift, ModAdd, ModMult                                                                            \\ \cline{2-7} 
                                                                                 & SM4~\cite{sm4_report}                & 32                    & 128                                                                      & 128                                                               & 8                                                                                    & XOR, S-Box, Shift                                                                                      \\ \cline{2-7} 
                                                                                 & Camellia~\cite{camellia}           & 18 / 24                & 128 / 192 / 256                                                          & 128                                                               & 8                                                                                    & XOR, S-Box, Shift, AND, OR                                                                             \\ \hline
\textbf{HASH}                                                                    & SHA-2~\cite{sha}              & 64                    & -                                                                        & 512                                                               & 32                                                                                   & XOR, NOT, AND, Shift, ModAdd                                                                        \\ \hline
\multirow{2}{*}{\textbf{PUK}}                                                    & RSA~\cite{rsa_first}                & 1                     & 1024 / 2048 / 3072 / 4096                                                & 256 / 512                                                            & $\geq$ 1024                                                                                   & AND, OR, Shift, ModMult, ModExp                                                                        \\ \cline{2-7} 
                                                                                 & ECDSA~\cite{ecdsa}              & 1                     & 160 / 224 / 256 / 384 / 521                                              & 256 / 512                                                            & $\geq$ 256                                                                                   & AND, OR, Shift, ModMult, ModInv                                                                \\ \hline
\multirow{3}{*}{\textbf{PQC}}                                                    & Dilithium~\cite{dilithium, dilithium_paper}          & 1                     & 2048 / 3072                                                              & 256 / 512                                                            & 23                                                                                   & XOR, AND, Shift, NTT, Keccak, ModMult, Comparison                                                               \\ \cline{2-7} 
                                                                                 & Falcon~\cite{falcon_first}             & 1                     & 896 / 1280                                                               & 256 / 512                                                            & 64                                                                                   & XOR, AND, Shift, FFT, ModMult, Comparison \\ \cline{2-7} 
                                                                                 & SPHINCS+~\cite{sphincs}           & 1                     & 256 / 384 / 512                                                          & 256 / 512                                                            & 32                                                                                   & XOR, AND, Shift, Keccak, Comparison                                                                    \\ \hline\hline
\end{tabular}
}
\\[1.0ex]
\begin{minipage}{0.94\textwidth}
\footnotesize
\textsuperscript{*} \textbf{Round} In PKC and PQC, the overall operation is not performed in a repetitive round-based manner, but rather consists of a single large mathematical operation, which is why the round count is set to 1. However, despite having a round count of 1, the internal operations may involve multiple rounds. For instance, in RSA, the modular exponentiation is performed $O(\log(e))$ times, depending on the length of the public key exponent $e$~\cite{rsa_first}, whereas in PQC, the Keccak-f1600 permutation function is executed 24 times~\cite{keccak_permutation}.\\
\textsuperscript{\dag} \textbf{Input Unit Size} In block ciphers and hash functions, the input unit size represents the block size processed in each operation. For public-key cryptography (PUK) and post-quantum cryptography (PQC), it corresponds to the length of the message digest, computed via widely used cryptographic hash functions such as SHA-256 or SHA-512.\\
\textsuperscript{$\ddagger$} \textbf{Processing Granularity} In PUK schemes, the processing granularity represents the bit length of the prime modulus, whereas in PQC, it corresponds to the bit width of polynomial coefficients. Although PQC generally operates on smaller bit-width units than PUK, it achieves comparable security levels through high-degree polynomial operations or hash-based constructions~\cite{pqc_small_length}.
\end{minipage}
\vspace{-2mm}
\end{table*}

%\vspace{+1mm}
\subsubsection{Latency of PKC and PQC Signing} Fig.~\ref{fig:moti_puk_sign}\label{sec:sining_motivation} presents the latency breakdown of digital signature generation using various PKC and PQC algorithms in a tamper-proof logging scenario, where signatures are generated at runtime for individual log entries.
Since logging operations typically target small messages ranging from a few hundred bytes to several kilobytes~\cite{tamper_logging_size}, the evaluation was conducted with input message sizes ranging from 256 B to 10 KB.
Given such small message sizes, PKC and PQC operations dominate the total latency.
%the cryptographic operations involved in PKC and PQC algorithms account for a significant portion of the total processing latency.
%Commercial tamper-proof logging systems, such as Amazon's EmLog and Microsoft’s HARDLOG, are required to complete log generation and signing for each event within 2 to 15 milliseconds to meet real-time processing requirements.
Commercial tamper-proof logging systems must complete log generation and signing for each event within 2 to 15 ms to meet real-time processing requirements~\cite{emlog_logging, tamper_logging_size}.
However, as shown in the figure, Falcon 1024 and SPHINCS+ require considerable time for signature generation regardless of the input message size.
%, making them impractical for latency-critical tamper-proof logging.}
%\samnotes{But, Dilithium and Falcon seems to meet the requirement, no? Generalizing too easily.}
%This indicates that performing PQC-based signing on the CPU is impractical for implementing tamper-proof logging systems that must adhere to such strict latency constraints.
This indicates that several NIST-standardized PQC algorithms are impractical for latency-critical tamper-proof logging when executed on the host CPU, as is typical in modern SSDs.

%\vspace{-0.5mm}
\section{Design Considerations}\label{sec:design_requirement}
% \vspace{-1mm}
%As discussed in Section~\ref{sec:motivation}, modern storage systems are increasingly required to support a wide range of cryptographic algorithms to address evolving security demands.
%However, current SSD hardware supports only a limited set of fixed algorithms, forcing most other cryptographic operations to rely on the host CPU, which in turn leads to performance degradation.
%To overcome these limitations, this work proposes an in-NAND self-encryption architecture that performs various cryptographic algorithms directly within the NAND flash chip.
%This section discusses the key design considerations for realizing such an in-NAND self-encryption architecture.
This section presents the key design considerations behind FlashVault, an in-NAND self-encryption architecture designed to support a wide range of cryptographic algorithms with low-latency execution.

%\subsection{Area and Power Constraints}\label{sec:resource_constraint}\vspace{-1mm}
%\vspace{-1mm}
\subsection{Available Silicon Area}\label{sec:area_constraint}
% \vspace{-1mm}
%A key contribution of the Flashvault design is the integration of cryptographic engines within the unused area of 4D NAND flash.
A key feature of FlashVault is integrating cryptographic engines into the unused area under the memory array of a 4D V-NAND chip.
%To enable this integration beneath the memory array, we first estimate the free space in 4D NAND flash.
To enable this integration in the empty space, we first estimate the available area in the 4D V-NAND.
%Since detailed information about this space has not been publicly disclosed, we investigate the area characteristics of Samsung’s 7\textsuperscript{th}-generation 4D V-NAND flash chip, which is the company’s first 4D V-NAND product~\cite{samsung_7}.
Due to lack of public data, we analyze Samsung’s first 4D V-NAND product, the 7\textsuperscript{th}-generation 4D V-NAND chip~\cite{samsung_7}.
We calculate the available space as \(M - P\), where \(M\) and \(P\) are the memory array and peripheral circuit areas.
%denote the areas of the memory array and the peripheral circuit, respectively.
To estimate the peripheral circuit area, we refer to the specifications of Samsung’s 6\textsuperscript{th}-generation 3D V-NAND solution~\cite{samsung_6}.
%, which precedes the 7\textsuperscript{th}-generation.
The 7\textsuperscript{th}-generation NAND flash chip improves area efficiency by repositioning the peripheral circuit underneath the memory array. 
As documented in~\cite{4d_nand_survey, pif, CUA_patent}, the peripheral circuit lies entirely under the memory array.
%the peripheral circuit in this design is entirely contained beneath the memory array.
% [Revision] We estimate the peripheral circuit area using the formula \(D_6 - D_7\), where \(D_6\) and \(D_7\) are the die sizes of the 6\textsuperscript{th} and 7\textsuperscript{th}-generation NAND flash chips, respectively.
Peripheral area is estimated as \(D_6 - D_7\), where \(D_6\) and \(D_7\) are die sizes of the 6\textsuperscript{th} and 7\textsuperscript{th}-genration chips.
Consequently, we identify a free space of 20.60 $mm^{2}$ in the 7\textsuperscript{th}-generation 4D V-NAND flash. 
%\textbf{\textit{(Feature 1)}} We \textit{leverage this unused area to integrate the cryptographic engines} for in-storage encryption and decryption.
\textbf{(Takeway 1)} \textit{We leverage this unused area (20.60 $mm^{2}$) to integrate cryptographic engines supporting diverse algorithms.}

%\vspace{-1mm}
\subsection{Power Budget}\label{sec:power_constraint}
% \vspace{-1mm}
Since a NAND flash chip receives only a limited amount of power from the main power management IC of the SSD~\cite{power_supply}, any cryptographic engine integrated into the unused space of 4D V-NAND must stay within this power budget.
%must operate within this constrained power budget.
If the proposed architecture exceeds this budget, it may degrade cell reliability or performance due to thermal issues~\cite{heatwatch, thermal_performance}.
Thus, we identify the power budget and ensure that cryptographic engines operate safely within this limit.
However, as Samsung’s 7\textsuperscript{th}-gen V-NAND power budget is undisclosed, we estimate it based on the power budget of the 6\textsuperscript{th}-generation and the reported efficiency improvements in the 7\textsuperscript{th}-generation.
Since block-level programming typically consumes the highest power among NAND flash operations~\cite{pif, power_dominant_program}, we use it as the basis for power budget estimation. 
For 6\textsuperscript{th}-generation V-NAND, the programming power can be calculated as follows: \vspace{-2mm}
\begin{equation}\vspace{-2mm}
P_{\text{program}} = I_{\text{program}} \cdot V_{\text{max}}, \vspace{-1mm}
\end{equation}
where $P_{\text{program}}$ denotes the programming power, $I_{\text{program}}$ is the programming current, and $V_{\text{max}}$ is the maximum operating voltage.
Based on technical documents~\cite{samsung_6, samng_current}, the programming current and maximum operating voltage of the 6\textsuperscript{th}-generation V-NAND are 13.8 mA and 3.6 V, respectively, resulting in a power budget of 49.68 mW.
The 7\textsuperscript{th}-generation V-NAND achieves 16\% improvement in power efficiency over the 6\textsuperscript{th} generation~\cite{power_efficiency_improve}; thus, its power budget is estimated to be 41.73 mW.
\textbf{(Takeway 2)} Given this estimated budget, \textit{we design the cryptographic engines to run in parallel under the power constraint (41.73 mW) within the 4D V-NAND die.}

%Given this estimated budget, \textit{the cryptographic engines are configured for parallel deployment within the 4D V-NAND die, ensuring that the power constraint (41.73 mW) is satisfied.}

\begin{figure}[t]
    \centering \includegraphics[scale=0.8]{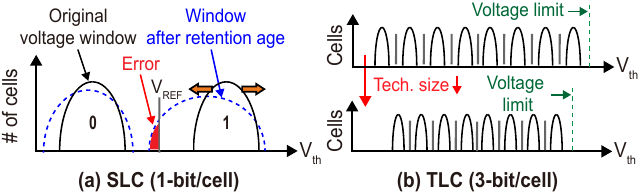}\vspace{-1mm} 
    \caption{Threshold voltage ($V_{th}$) distributions of (a) single-level cell (SLC) and (b) triple-level cell (TLC).}  \label{fig:design_challenge_retention}
    \vspace{-3mm}
\end{figure}

\begin{figure*}[t]
    \centering
    \includegraphics[scale=0.59]{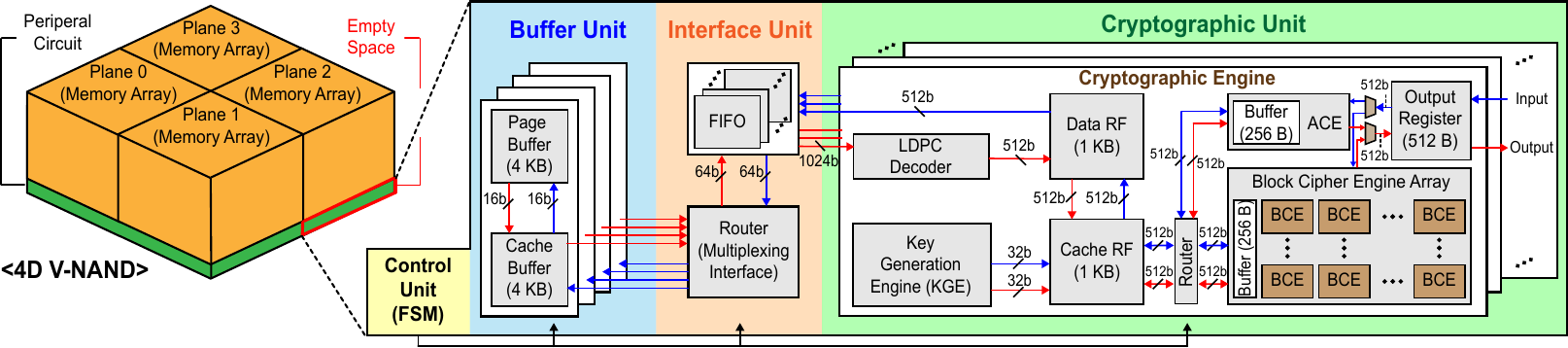} \vspace{-1mm}
    \caption{Overview of the FlashVault architecture. Red and blue arrows indicate the dataflows for read and program operations, respectively. BCE and ACE represent the block cipher engine and the asymmetric cipher engine (supporting PKC and PQC).}  
    \label{fig:flashvault_architecture}
    \vspace{-3mm}
\end{figure*}

%\vspace{-1mm}
\subsection{Reconfigurable Cryptographic Engine}\label{sec:challenge_crypto}
% \vspace{-1mm}
%One of the key design considerations of FlashVault is the versatility of its cryptographic engine to support a wide range of cryptographic algorithms.
One key design goal of FlashVault is supporting diverse cryptographic algorithms through a versatile cryptographic engine.
%To identify the fundamental cryptographic primitives used across different schemes, we thoroughly analyzed 17 block ciphers provided by the Botan library~\cite{botan}, public-key cryptosystems based on integer factorization and elliptic curve problems, and PQC algorithms, including lattice-based and hash-based schemes, available in the liboqs library~\cite{liboqs}.
To identify the fundamental cryptographic primitives used across different schemes, we analyzed 17 block ciphers from the Botan library~\cite{botan}; public-key cryptosystems based on integer factorization and elliptic curve problems; and PQC algorithms, including lattice-based and hash-based schemes, from the liboqs library~\cite{liboqs}.
Table~\ref{tab:cipher_operation_summary} summarizes the key primitive operations used in widely adopted block ciphers and public-key algorithms, and NIST-standardized PQC schemes~\cite{nist_pqc_altorithms}.
Our analysis reveals that block ciphers are typically composed of five basic operations: modular arithmetic, logical operations, permutation, substitution (S-Box), and bitwise shift.
PKC and PQC algorithms, on the other hand, primarily rely on logical operations, bitwise shifts, number-theoretic transforms, Keccak, modular arithmetic, and comparison operations.
%NTT/FFT
\textbf{(Takeway 3)} Based on these observations, \textit{we develop a reconfigurable cryptographic engine that supports diverse primitives, enabling FlashVault to efficiently implement various algorithms within the NAND die.}

%\textit{Based on these observations, we designed a reconfigurable cryptographic engine capable of flexibly supporting diverse cryptographic primitives, thereby enabling FlashVault to efficiently implement a wide range of cryptographic algorithms within the in-NAND die.}

\begin{figure*}[t]
    \centering
    \includegraphics[scale=0.54]{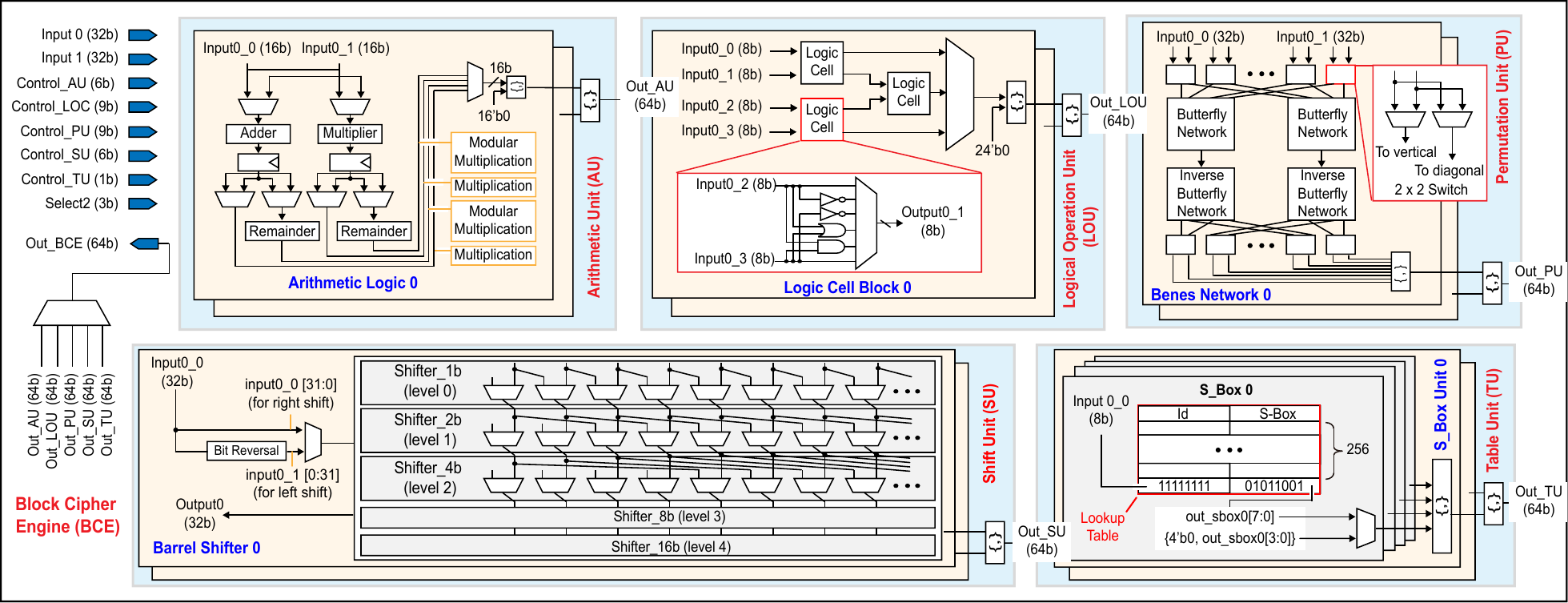}\vspace{-2mm}
    \caption{Schematic diagram of the Block Cipher Engine (BCE). Each BCE has five fundamental building blocks: arithmetic unit (AU), logical operation unit (LOU), permutation unit (PU), shift unit (SU), and table unit (TU).}  
    \label{fig:bce}
    \vspace{-4mm}
\end{figure*}

%\vspace{-1mm}
\subsection{Error Correction}\label{sec:consideration_ecc}
% \vspace{-1mm}
NAND technology scaling has evolved beyond physical scaling, transitioning from single-level cell (SLC) to triple-level cell (TLC) designs.  
While multi-bit memory cells significantly increase storage density and reduce cost per bit, this transformation has deteriorated the reliability characteristics of NAND flash, particularly in terms of data retention and endurance.
Fig.~\ref{fig:design_challenge_retention}-(a) shows the threshold voltage ($V_{\text{th}}$) distributions of SLC and how they are affected by various error sources.  
According to previous studies~\cite{modern_ssd_asplos, evanesco, heatwatch, nand_pomacs, 3d_nand_reliability_nand}, reading or programming a flash cell slightly increases the $V_{\text{th}}$ of adjacent cells by unintentionally injecting electrons into their charge traps (i.e., cell-to-cell interference).  
In addition, the leakage of electron charges over time decreases the $V_{\text{th}}$ level (i.e., retention loss).  
As a result of these mechanisms, the $V_{\text{th}}$ level can drift across the reference voltage ($V_{\text{REF}}$), leading to bit errors.
As shown in Fig.~\ref{fig:design_challenge_retention}-(b), TLC has narrower voltage windows for each state than SLC, and the technology scaling further shrinks the size of these windows.  
To ensure data reliability under such conditions, modern SSDs incorporate error correction codes (ECC) within their controllers.
%~\footnote{To prevent confusion, we denote Elliptic-Curve Cryptography as `ECC' and Error Correction Codes as `ECCs' in this paper.}.
Since FlashVault performs cryptographic operations directly within the NAND chip, it is essential to correct bit errors in advance  to ensure only error-free data is encrypted. 
\textbf{(Takeway 4)} To support reliable in-NAND self-encryption, \textit{FlashVault integrates error correction capabilities into the NAND flash chip, allowing correction to take place on-chip.}

%\vspace{-1mm}
\section{FlashVault Architecture}\label{sec:overall_architecture}
% \vspace{-1mm}
In this section, we present an overview of the proposed on-die self-encryption architecture, FlashVault.  
%FlashVault enhances data security, performance, and energy efficiency by addressing the limitations of CPU-offloading-based encryption systems.  
FlashVault integrates cryptographic engines into the unused area of a NAND flash chip.
%, utilizing die space that has not been exploited in prior 4D V-NAND SSD architectures.
This strategic integration utilizes idle silicon, distinguishing FlashVault from state-of-the-art SSDs.
%represents an effective utilization of idle silicon area, distinguishing FlashVault from state-of-the-art SSDs.
Fig.~\ref{fig:flashvault_architecture} illustrates the detailed architecture of FlashVault, comprising (i) a control unit, (ii) a buffer unit, (iii) an interface unit, and (iv) a cryptographic unit embedded in the unused die area.

\noindent\textbf{1) Control Unit:}
FlashVault includes a control unit based on a finite-state machine (FSM), which operates as a supplementary unit to the SSD controller. 
It receives control signals issued by the SSD controller firmware in response to host I/O commands.
Based on these signals, the FSM manages the control flow and selects the appropriate hardware resources and dataflow paths for each state.

\noindent\textbf{2) Buffer Unit:} The buffer unit comprises a 4 KB page and cache buffers.
The page buffer holds data directly sensed from the vertically stacked memory array, while the cache buffer mirrors this data. 
This double-buffering scheme enhances I/O performance by allowing the cache buffer to forward previously transferred data to downstream modules in parallel with ongoing read or program operations in the page buffer.

\noindent\textbf{3) Interface Unit:} The interface unit arbitrates plane access to the cryptographic units through an internal router. It temporarily buffers incoming data from the buffer unit, aligns it, and forwards it to the cryptographic unit via the input bus.

\noindent\textbf{4) Cryptographic Unit:} The core component of FlashVault is the cryptographic unit, which consists of multiple cryptographic engines. 
Each engine is configured as an array of \textit{block cipher engines (BCEs)} for symmetric encryption and an \textit{asymmetric cipher engine (ACE)} supporting PKC and PQC~\footnote{Block ciphers use the same secret key for both encryption and decryption, whereas PKC and PQC use asymmetric key pairs comprising a public key for encryption and a private key for decryption.}.
When FlashVault operates in self-encrypting mode, block ciphers are executed for every I/O request.
These algorithms allow subword-level parallelism across multiple processing stages (i.e., each round in the cipher)~\cite{block_cipher_review}.
Thus, FlashVault adopts a SIMD-style array of 16 BCEs, enabling high throughput.
The ACE module is activated during signature generation and verification.
Since most cryptographic operations in PKC and PQC, excluding hash computations, are performed sequentially~\cite{survey_pkc}, FlashVault incorporates a single ACE module.

In addition to the cipher engines, the cryptographic engine integrates auxiliary components to support full in-NAND cryptographic functionality.
Among them, a \textit{low-density parity-check (LDPC) decoder} performs error correction on data read from the NAND array.
%serves as an essential module for error correction on data read from the NAND array. 
Unlike conventional SSDs, which perform error correction in the SSD controller, FlashVault conducts in-situ correction within the NAND flash chip. 
For this purpose, the LDPC decoder first corrects raw data errors from the NAND array and then forwards the decoded data to two dual-purpose register files: the Data Register File (DRF) and the Cache Register File (CRF). 
FlashVault adopts the LDPC decoder in~\cite{ldpc}, which uses a gradient descent bit-flipping algorithm. 
%It initializes all bits (i.e., variable nodes) using a hard decision, which means interpreting the signal as 0 or 1.
It initializes all bits (i.e., variable nodes) using a hard decision, which interprets the signal as 0 or 1.
The decoder then iteratively flips the bit with the lowest reliability score, determined from the parity check results and NAND read channel conditions.
This process continues until all parity constraints are satisfied or a predefined iteration threshold is reached.
%Despite being based on a hard-decision architecture with a rate-0.88 quasi-cyclic LDPC code, the decoder achieves error correction performance comparable to that of soft-decision decoders. 
Despite being based on a hard-decision architecture with a rate-0.88 quasi-cyclic LDPC code, the decoder provides error correction performance 
%to soft-decision decoders, which interpret signals probabilistically rather than using binary thresholds.
comparable to soft-decision decoders that use probabilistic signal interpretation.

As another auxiliary component, FlashVault includes a Physical Unclonable Function (PUF)-based 
%\samnotes{What is PUF?}
\textit{key management engine (KME)}.
This engine integrates a ring oscillator (RO) PUF that leverages process-induced variations in oscillator frequencies to generate a chip-unique bit pattern, which is used as the root key~\cite{ro_puf}.
The root key is then expanded by a key derivation function (KDF) into symmetric session keys and asymmetric key pairs.
%symmetric session keys and asymmetric key pairs for cryptographic use.
The KDF uses a secure hash function, combining the root key with additional salt (i.e., a random value) or context, producing keys with varying lengths and formats, tailored to meet security requirements.
The derived keys are stored in the CRF and delivered to either the BCEs or the ACE via the internal router.

%\vspace{-1mm}
\section{Cipher Engine Microarchitecture}\label{sec:Microarchtiecture}
% \vspace{-1mm}
%One of the key features of FlashVault is its ability to support a wide range of cryptographic algorithms. 
This section describes the core modules of FlashVault
% , i.e., BCE and ACE, 
building on the algorithmic analysis discussed in Section~\ref{sec:challenge_crypto}.

%\vspace{-1mm}
\subsection{Block Cipher Engine (BCE)}\label{sec:bce_structure}
% \vspace{-1mm}
As discussed in Section~\ref{sec:challenge_crypto}, we identify five fundamental operations commonly used in block cipher algorithms: modular arithmetic, logical, permutation, S-Box, and shift operations.
Accordingly, we implement five corresponding compute units, that is, arithmetic, logical operation, permutation, shift, and table, in the BCE.
%\samnotes{revised: please check}
%%Accordingly, we implement five corresponding compute units in the BCE, which are an arithmetic unit, a logical operation unit, a permutation unit, a shift unit, and a table unit.
%These units are designed with a reconfigurable structure to support various block cipher operations. 
These units are reconfigurable to support various block cipher operations. 
Fig.~\ref{fig:bce} shows the microarchitecture of the BCE, which integrates these five units.

\noindent\textbf{1) Arithmetic Unit (AU):} Some block cipher algorithms require modular addition and/or multiplication. 
To support these operations efficiently, the AU features dedicated hardware components: an adder, a multiplier, and two remainder circuits. 
As illustrated in Fig.~\ref{fig:bce}, the outputs of the adder and multiplier can be routed either to the remainder circuits for modular operations or directly to the output for non-modular operations.

%FlashVault performs modular operations with fewer cycles than the embedded processors used in conventional SSDs, owing to its specialized design for modular arithmetic.
%For example, while an ARM Cortex-M23 processor takes 17 or 34 cycles to perform a single modulo operation\cite{arm_cpu}, the AU of FlashVault accomplishes the same task in just 2 to 4 cycles.

\noindent\textbf{2) Logical Operation Unit (LOU):} Block cipher algorithms use the same key for encryption and decryption, performing these operations through bitwise XOR between the key and input data. 
%The LOU supports XOR and extends its functionality to additional logical operations such as OR, AND, NOT, and combinations of these logical primitives.
The LOU supports XOR and extends to additional operations such as OR, AND, NOT, and their combinations.

\noindent\textbf{3) Permutation Unit (PU):} Block cipher algorithms utilize permutation operations to obscure the relationship between the input (i.e., plaintext) and the output (i.e., ciphertext) during encryption. 
The PU is equipped with two 32-bit Benes networks to support this. 
These networks can either be combined to produce a single 64-bit permutation or operate independently to generate two separate 32-bit permutation results.

\noindent\textbf{4) Shift Unit (SU):} Some block ciphers use shift operations to increase the complexity of deducing key values from encrypted data.
For example, in AES, each row of a 4$\times$4 data block is shifted by a different offset to shuffle data locations~\cite{aes_shift}.
The SU supports such operations and consists of two barrel shifters.
These shifters can be combined to support up to 64-bit arithmetic, logical, and rotate shift operations.
%To support both left and right shift directions, the input data can either be passed through directly or bit-reversed.

\noindent\textbf{5) Table Unit (TU):} To obscure the statistical properties of the plaintext, the S-Box maps input bits to corresponding output bits using a predefined substitution table.
We implement the S-Box operation using lookup tables.
%Specifically, the TU includes four S-Box units, each consisting of a lookup table with 256 byte-wide entries.
Specifically, the TU includes four S-Box units, each with a 256-entry byte-wide lookup table.
These tables primarily convert 8-bit input data into 8-bit output values.
%For data with wider or narrower lengths, multiple tables can be combined to function as a single S-Box, or a single table can perform a substitution with zero-padding at the most significant bits for narrower data. 
For wider input data, multiple tables can be combined to function as a single S-Box.
For narrower input data, a single table can perform substitution with zero-padding on the most significant bits.
% [revision] In our proposed architecture, tables for various S-Box operations of block cipher algorithms are stored in the DRAM of SSDs.
% [revision] When the encryption or decryption of a specific algorithm is triggered, the corresponding table is transferred to the register file within the flash chip.

\begin{figure}[t]
    \centering
    \includegraphics[scale=0.65]{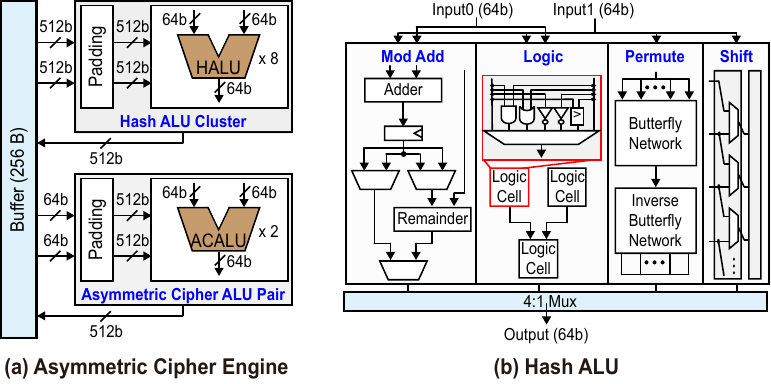}\vspace{-1mm} 
    \caption{(a) Architectural schematic of the asymmetric cryptographic engine (ACE), consisting of a hash ALU cluster and a pair of asymmetric cipher ALUs. (b) Internal structure of the hash ALU.}  
    \label{fig:ace_diagram}
    \vspace{-4mm}
\end{figure}

%\vspace{-1mm}
\subsection{Asymmetric Cryptographic Engine (ACE)}\label{sec:ace_structure}
% \vspace{-1mm}
Fig.~\ref{fig:ace_diagram}-(a) depicts the structure of ACE, designed for digital signature generation and verification. 
ACE integrates a Hash ALU cluster for hash computations and two asymmetric cipher ALUs for public-key and post-quantum cryptography.

\vspace{1mm}
\noindent\textbf{1) Hash ALU Cluster:} Digital signature algorithms typically hash messages, then sign the resulting digest~\cite{nist_hash}.
%\samnotes{?? looks like an incomplete sentence}
%Digital signature algorithms typically do not sign the original message directly; instead, they compute a hash of the message and apply the signature operation to the resulting digest~\cite{nist_hash}.
%Hashing enables the signature process to efficiently handle messages of arbitrary length by reducing them to a fixed-size digest.
Hashing enables efficient handling of arbitrary-length messages by reducing them to fixed-size digests.
Furthermore, certain PQC schemes like SPHINCS+~\cite{sphincs} rely on hash functions as core primitives.
%To support both the hashing stage of conventional signature generation and the core operations of hash-based PQC algorithms, the ACE integrates a dedicated Hash ALU cluster.
To support conventional signature hashing and hash-based PQC, ACE integrates a dedicated Hash ALU cluster.
%Since hash functions can process input by dividing it into fixed-size blocks and can operate on multiple blocks independently via tree hashing, they are inherently parallelizable~\cite{hash_parallel_survey}.
Hash functions are inherently parallelizable [6] as they process input by dividing it into fixed-size blocks and operating independently via tree hashing~\cite{hash_parallel_survey}.
To support this parallelism, FlashVault incorporates eight Hash ALUs to accelerate the hashing process.

Fig.~\ref{fig:ace_diagram}-(b) shows the block diagram of the Hash ALU, which comprises four key components optimized for the primitive operations used in cryptographic hash functions: (i) a modular addition unit, (ii) a logic operation unit, (iii) a permutation unit, and (iv) a shift unit. 
The \textit{modular addition unit} supports both standard and modular additions for hash state updates.
%both standard and modular additions to update the internal state of the hash computation.
The \textit{logic operation unit} supports bitwise operations such as XOR, AND, and NOT to introduce non-linearity by disrupting linear relationships between input and output.
%The \textit{permutation unit}, implemented using a Benes network, globally reorders bit positions to promote data diffusion and reduce structural correlations.
The  \textit{permutation unit}, based on a Benes network, reorders bit positions to enhance diffusion.
The \textit{shift unit} supports word-level (e.g., 32- or 64-bit) logical and circular shifts, as commonly used in hash functions like SHA-2~\cite{sha}. 

\noindent\textbf{2) Asymmetric Cipher ALUs:} PKC and PQC computations typically proceed sequentially, with each stage depending on prior results, e.g., modular exponentiation, scalar multiplication, and NTT/FFT-based transforms.
%as each stage depends on the result of the previous one (e.g., modular exponentiation, scalar multiplication, and NTT/FFT-based transforms).  
To efficiently support this, ACE integrates two asymmetric cipher ALUs, which include the arithmetic, logic, permutation, and shift units. 
%\samnotes{revised: please check.}
%%To efficiently support this, ACE integrates two asymmetric cipher ALUs. Each asymmetric cipher ALU include four units: an arithmetic unit, a logic unit, a permutation unit, and a shift unit. 
%These units are designed to handle a wide range of high-precision integer operations, bit manipulations, and structural data transformations.
These units handle high-precision integer arithmetic, bit manipulations, and data transformations.
%While the constituent units are architecturally similar to those in the Hash ALU, the asymmetric cipher ALUs are designed to support more diverse and complex arithmetic operations required by PKC and PQC computations.
While architecturally similar to the Hash ALU, asymmetric cipher ALUs support more diverse and complex arithmetic for PKC/PQC. 
%, such as modular arithmetic and large-integer processing.
That is, the \textit{arithmetic unit} includes extended functional blocks beyond the modular addition unit of the Hash ALU, including adders, multipliers, and modular operators.
%The arithmetic unit adopts a fixed 64-bit integer datapath. 
%Narrower operands are zero-extended for datapath alignment.
The arithmetic unit adopts a fixed 64-bit integer datapath, zero-extending narrower operands for alignment.
%Operands narrower than 64 bits are zero-extended to align with the datapath width. 
%For high-precision computations, each operand is divided into 64-bit limbs (i.e., subwords), and the computation is carried out in a limb-wise manner, with coordinated carry propagation and cross-limb alignment.
For high-precision computations, operands are divided into 64-bit limbs (subwords), with limb-wise computation involving carry propagation and cross-limb alignment.
The adder in the arithmetic unit is designed with a carry-save structure to efficiently support multi-limb accumulation.
%The multiplier and modular operator are implemented to achieve low propagation delay, employing a Wallace tree structure~\cite{wallace_multiplier} and a Barrett reduction scheme~\cite{barrett_reduction} using shifters and fixed-coefficient multipliers, respectively.
To reduce propagation delay, the multiplier uses a Wallace tree~\cite{wallace_multiplier} and the modular unit uses Barrett reduction~\cite{barrett_reduction} with shifters and fixed-coefficient multipliers.

%\vspace{-1mm}
\section{Algorithm Selection and Reconfiguration}\label{sec:argorithm_change}
% \vspace{-1mm}
% FlashVault flexibly supports the execution of various cryptographic algorithms.
To select a cryptographic algorithm in FlashVault, a user-level application issues an \texttt{ioctl()} system call with the desired algorithm identifier as an argument.
This transfers an NVMe command containing the feature data of the selected algorithm to the system buffer in the kernel space.
%When the \texttt{ioctl()} function is invoked to switch cryptographic algorithms, it transfers an NVMe command structure containing the feature data of the selected algorithm to the system buffer in the kernel space. 
The device driver then issues an \texttt{nvme\_set\_feature} command, which updates control registers in the SSD controller.
These registers propagate a control signal to the FSM unit in FlashVault, which interprets the selected algorithm and reconfigures the cryptographic pipeline by enabling the necessary modules and associated datapaths.
This hardware-level reconfiguration is performed dynamically at runtime, enabling FlashVault to support diverse cryptographic algorithms without requiring firmware updates or system reboot.

%\vspace{-1mm}
\section{Evaluation}\label{sec:evaluation}

% \vspace{-1mm}
\subsection{Area and Power Analysis of FlashVault}\label{sec:area_analysis}
% \vspace{-1mm}
\noindent \textbf{1) Methodology:} %To quantitatively evaluate the effectiveness of FlashVault, we implemented all components shown in Fig.~\ref{fig:flashvault_architecture} at the register-transfer level (RTL) using Verilog.
%To verify functional correctness, we executed each cryptographic algorithm in software to generate golden reference values.
%Intermediate states and final results were extracted by instrumenting the software model and compared with RTL outputs using a cocotb-based testbench~\cite{cocotb}.
%For area and power analysis, we synthesized the design and performed place-and-route (P\&R) using Synopsys EDA tools at fast/fast and slow/slow corners in a 65 nm CMOS process.
To evaluate the effectiveness of FlashVault, we synthesized all components shown in Fig.~\ref{fig:flashvault_architecture} and performed place-and-route (P\&R) using Synopsys EDA tools at fast/fast and slow/slow corners in a 65nm CMOS process.
For low-power design, we utilized the unified power format (UPF) to implement power gating. 
%with required isolation cells and retention registers.
Additionally, for dynamic power minimization, we implemented clock gating at the RTL level.
%To balance performance and leakage power, we selectively used standard cell libraries with different threshold voltages. 
%Low-$V_{\text{th}}$ cells were applied to timing-critical paths to meet performance targets, while high-$V_{\text{th}}$ cells were used in non-critical regions to minimize leakage power.
% [revision] To balance performance and leakage power, we selectively used standard cell libraries with different threshold voltages: applying Low-$V_{\text{th}}$ cells to timing-critical paths, and high-$V_{\text{th}}$ cells in non-critical regions.
On-chip memories were instantiated using commercial memory IPs.
Area was extracted using Synopsys IC Compiler II~\cite{synopsys_icc} and consumed power was estimated using PrimeTime PX~\cite{synopsys_pt} based on switching activity interchange format (SAIF) generated from post-layout simulations with StarRC-extracted parasitics~\cite{rc_synopsys}.
%including parasitics extracted by StarRC~\cite{rc_synopsys}.
To report results under more advanced technology assumptions, we scaled the area and power consumption to 14nm technology node using scaling factors from~\cite{scale_factor}.
This node selection reflects the technology used in recent commercial SSDs, which commonly adopt nodes in the mid-10nm range, i.e., 14nm~\cite{14nm_nand, 4d_nand_survey, nand_roadmap}.

\begin{table}[t]
\centering
\caption{Area and power of the hardware components in FlashVault}%\vspace{-2mm}
\vspace{-2mm}\
\label{tab:pnr_result}
\scalebox{0.75}{
\begin{tabular}{cccc}
\hline\hline
\multicolumn{4}{c}{\textbf{Block Cipher Engine (Bottom Level)}}                                                                                                                                                                                                                                                                     \\ \hline
\textbf{Component}                                                               & \textbf{Configuration}                                                                                                                                                                           & \textbf{Area {[}$\mu m^{2}${]}} & \textbf{Power {[}$mW${]}} \\ \hline
Arithmetic Unit                                                                  & 2 $\times$ Arithmetic Logic                                                                                                                                                                             & 1,263.15                & 0.10                  \\ \hline
Logical Operation Unit                                                           & 2 $\times$ Logic Cell Block                                                                                                                                                                             & 99.93                & 0.02                  \\ \hline
Permutation Unit                                                                 & 2 $\times$ Benes Network                                                                                                                                                                                & 247.19                & 0.08                  \\ \hline
Shift Unit                                                                       & 2 $\times$ Barrel Shifter                                                                                                                                                                               & 205.83                & 0.05                  \\ \hline
Table Unit                                                                       & 2 $\times$ S-Box Unit                                                                                                                                                                                   & 2,195.22               & 1.17                  \\ \hline
\textbf{Total}                                                                   &                                                                                                                                                                                                  & \textbf{4,2616.21}      & \textbf{0.41}         \\ \hline\hline
\multicolumn{4}{c}{\textbf{Asymmetric Cipher Engine (Bottom Level)}} \\ \hline
\textbf{Component} & \textbf{Configuration} & \textbf{Area [$\mu m^{2}$]} & \textbf{Power {[}$mW${]}} \\ \hline

Hash ALU Cluster & 
\begin{tabular}[c]{@{}c@{}}8 $\times$ Hash ALU,\\ 1 Padding Unit\end{tabular} & 
\begin{tabular}[c]{@{}c@{}}38,354.59\\ 1,491.57\end{tabular} & 
\begin{tabular}[c]{@{}c@{}}3.43\\ 0.06\end{tabular} \\ \hline

\begin{tabular}[c]{@{}c@{}}Asymmetric Cipher \\ ALU Pair (2 ACALUs)\end{tabular} & 
\begin{tabular}[c]{@{}c@{}}2 $\times$ ACALU,\\ 1 Padding Unit\end{tabular} & 
\begin{tabular}[c]{@{}c@{}}13,540.39\\ 1,605.35\end{tabular} & 
\begin{tabular}[c]{@{}c@{}}1.65\\ 0.08\end{tabular} \\ \hline

\textbf{Total} & & \textbf{54,991.89} & \textbf{5.13} \\ \hline\hline
\multicolumn{4}{c}{\textbf{FlashVault (Top Level)}}                                                                                                                                                                                                                                                                                 \\ \hline
\textbf{Component}                                                               & \textbf{Configuration}                                                                                                                                                                           & \textbf{Area [}$\mu m^{2}${]} & \textbf{Power {[}$mW${]}} \\ \hline
Control Unit                                                                     & 1 FSM                                                                                                                                                                                            & 399.24                & 0.09                  \\ \hline
Buffer Unit                                                                      & \begin{tabular}[c]{@{}c@{}}1 Page Bufffer (4 KB),\\ 1 Cache Buffer (4 KB)\end{tabular}                                                                                                           & 9,024.34              & 1.12                     \\ \hline
Interface Unit                                                                   & 1 FIFO, 1 Router                                                                                                                                                                                 & 2,538.33               & 2.31                \\ \hline
\multirow{6}{*}{Cryptographic Engine} 
& 1 LDPC Decoder & 58,624.29 & 6.65 \\
& 1 Key Generation Unit & 2,321.01 & 0.21 \\
& 2 $\times$ Registers (1 KB) & 2,654.89 & 0.19 \\
& 1 Output Register (512 B) & 1624.79 & 0.13 \\
& 1 Block Cipher Engine Array & 64,180.91 & 6.63 \\
& 1 Asymmetric Cipher Engine & 54,991.85 & 5.13 \\ \hline
\textbf{Total} & & \textbf{54,991.89} & \textbf{20.71} \\ \hline\hline
\end{tabular}
}
\vspace{-3mm}
\end{table}

\noindent\textbf{2) Analysis with Synthesized Results:} %As discussed in Section~\ref{sec:design_configuration}, placing cryptographic engines with integrated error correction modules in parallel at the die level improves I/O performance. 
%This section quantifies the maximum number of cryptographic engines that can be integrated within the given area and power budgets, based on P\&R results.
Table~\ref{tab:pnr_result} summarizes the post-layout area and power of each submodule in FlashVault. 
The control unit, buffer unit, and interface unit occupy 399.24, 9,024.34, and 2,538.33 $\mu m^{2}$, respectively, and consume 0.09, 1.12, and 2.31 $mW$ of power. 
%Additionally, the cryptographic engine occupies a total area of 0.18 $m {m}^2$ and consumes 18.87 $mW$ of power.
In addition, the cryptographic engine occupies 0.18 $m{m}^2$ and consumes 18.87 $mW$.
%Based on these results, we calculate the maximum number of cryptographic engines that can be integrated in parallel within a 4D V-NAND chip.
Placing cryptographic engines in parallel at the die level enhances I/O throughput.
Accordingly, we estimate the maximum number of engines ($N$) that can be integrated within a 4D V-NAND chip as follow:
%\samnotes{revised: please check}
%%Accordingly, we estimate the number of engines that can be integrated within a 4D V-NAND chip. The maximum number of engines ($N$) is calculated as:
%\vspace{-1mm}
\begin{equation}%\vspace{-1mm}
N = \min\left( \left\lfloor \frac{A_\text{budget}}{A_\text{engine}} \right\rfloor,\ \left\lfloor \frac{P_\text{budget}}{P_\text{engine}} \right\rfloor \right),
\end{equation}
where $A_\text{budget}$ and $P_\text{budget}$ denote the area and power budgets, while $A_\text{engine}$ and $P_\text{engine}$ represent the area and power of a single cryptographic engine.
By using post-layout results and budgets from Section~\ref{sec:area_constraint} and~\ref{sec:power_constraint}, we find that up to two cryptographic engines can be integrated in parallel. 
This translates to 32 BCEs, 16 Hash ALUs, and 4 asymmetric cipher ALUs that can be integrated within the unused area of a 4D V-NAND chip.
% As each cryptographic engine includes an array of BCEs, eight Hash ALUs, and two asymmetric cipher ALUs, up to 32 BCEs, 16 Hash ALUs, and 4 asymmetric cipher ALUs can be integrated within the unused area of a 4D V-NAND chip.

%\vspace{-1mm}
\subsection{Experimental Setup for Architectural Evaluation}\label{sec:archi_evaluation_method}
% \vspace{-1mm}
%\subsubsection{FlashVault Implementation and Baseline Systems}\label{sec:archi_evaluation_methodology}\vspace{-1mm}
\noindent \textbf{1) FlashVault Implementation:} As described in Section~\ref{sec:area_analysis}, FlashVault integrates two cryptographic engines within the area and power budgets of Samsung’s 7th-generation V-NAND. 
Based on this configuration, we evaluate the in-die cryptographic performance.
%the performance of FlashVault for in-die cryptographic operations. 
FlashVault is implemented using SimpleSSD~\cite{modern_ssd_amber}, incorporating the P\&R results in Table~\ref{tab:pnr_result} and the system parameters in Table~\ref{tab:ssd_configuration}.

\noindent\textbf{2) Baseline Systems:} We compare FlashVault with two baselines: (i) a Near-Core Processing (NCP) architecture and (ii) a real-machine system (CPU + SSD). 
The \textit{NCP architecture} uses the same cryptographic engines as FlashVault with identical configuration, but places them near the SSD controller instead of inside the NAND die. 
Following recent NCP designs~\cite{optimstore, ecssd, flagger, hardi_ssd_isca}, our NCP baseline uses a device DRAM as working memory for encryption.
The real-machine system performs encryption on the host CPU which follows the hardware setup provided in Section~\ref{sec:moti_exp_setup}.

\noindent\textbf{3) Methodology:}%\label{sec:archi_evaluation_method}
%To model the on-die processing behavior of FlashVault, we modified the hardware abstraction layers of SimpleSSD~\cite{modern_ssd_amber}.
%, including the I/O interface and NAND transaction paths, to incorporate the latency of cryptographic operations.
%The execution time of the cryptographic module was extracted from our custom cycle-approximate simulator and injected into the I/O request handling path, allowing the encryption overhead to be reflected in the overall SSD latency.
%The execution time of the cryptographic engine was estimated using a cycle-approximate simulator built upon design-for-test (DFT) verification results from Synopsys tools.
%The estimated cryptographic latency was injected into the I/O request handling path.
%, allowing the encryption overhead to be reflected in the overall SSD latency.
%For the NCP architecture, which performs cryptographic operations near the SSD controller, we inserted the corresponding latency into the controller-side I/O processing path within the abstraction layer to ensure that the overhead is properly accounted for in the system-level latency.
We modified the hardware abstraction layer of SimpleSSD~\cite{modern_ssd_amber} to enable end-to-end latency estimation for NVMe requests.
To reflect the latency overhead of cryptographic operations in FlashVault and NCP, we injected cycle-level latency estimated from design-for-test (DFT) verification using Synopsys tools into the I/O request handling path and controller-side I/O processing path of SimpleSSD, respectively.
%For the NCP architecture,we inserted the corresponding latency into the controller-side I/O processing path.
%to ensure that the overhead is properly accounted for in the system-level latency.
%Furthermore, we modified the DRAM abstraction layer to emulate read/write delays during cryptographic operations, based on timing parameters specified in Micron's LPDDR3 datasheet~\cite{lpddr3}.
In addition, we modified the DRAM abstraction layer in the NCP model to reflect the DRAM access latency during cryptographic operations.
%To enable end-to-end latency analysis, we added timing measurement points along the I/O processing path, allowing the simulator to report the overall latency experienced by NVMe requests.
The real-machine system follows the methodology described in Section~\ref{sec:moti_exp_setup}, with SSD latency derived from baseline SimpleSSD (without cryptographic engines) for fair comparison.
Our evaluation uses the block ciphers, PKC, and PQC algorithms listed in Section~\ref{sec:moti_nand_sed} as benchmarks, assessed by varying the input sizes.
%presented in Table~\ref{tab:cipher_operation_summary} as benchmarks.
%, assessed across various input sizes.

\begin{table}[t]
\centering
\caption{Simulated SSD configurations}\vspace{-2mm}
%\vspace{-2mm}
\label{tab:ssd_configuration}
\scalebox{0.88}{
\begin{tabular}{cc}
\hline\hline
\textbf{Component} & \textbf{Configuration}                                                                                                                                             \\ \hline\hline
Controller         & \begin{tabular}[c]{@{}c@{}}Arm Cortex-R8 Processor (@ 1.5 GHz)~\cite{Arm_cortex_r8}\\ 16 Gb LPDDR3~\cite{lpddr3}\end{tabular}                                                                         \\ \hline
Flash Memory       & \begin{tabular}[c]{@{}c@{}}4 Channels, 4 Packages, 2 Dies, 4 Planes\\ 682 Blocks, 128 Pages, Page Size = 4 KB\\ NV-DDR3, 1600 MT/s, 8-bit Channel Bus\end{tabular} \\ \hline
Cryptographic Unit & \begin{tabular}[c]{@{}c@{}} 2 Cryptographic Engines (@ 200 MHz)\\ (32 BCEs, 16 Hash ALUs, and 4 Asymmetric \\ {Cipher Engines)}\end{tabular} \\ \hline
Timing Parmeters   & \begin{tabular}[c]{@{}c@{}}t$_{R}$ = 45 $\mu$s, t$_{PROG}$ = 400 $\mu$s, t$_{ERASE}$ = 2 ms\\ t$_{PCBSY}$, t$_{RCBSY}$ = 3 $\mu$s,\\ Storage Stack Latency = 5 $\mu$s\end{tabular}                          \\ \hline\hline
%Energy Parameters   & \begin{tabular}[c]{@{}c@{}}Operation Voltage = 3.6 V\\ I$_{READ}$, I$_{PROG}$, I$_{ERASE}$ = 25 mA\\ I$_{BUSIDLE}$ = 5 mA, I$_{STDBY}$ = 10 uA\end{tabular}                                 \\ \hline\hline
\end{tabular}
}
\vspace{-3mm}
\end{table}

\begin{figure}[t]
    \centering
    \includegraphics[scale=0.74]{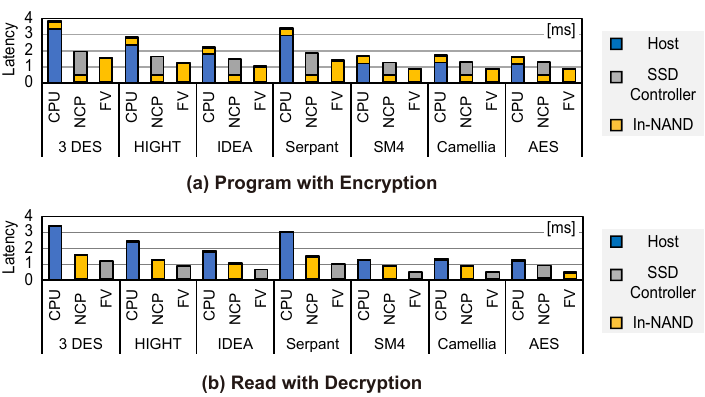}\vspace{-2mm} 
    \caption{(a) Program and (b) read latency breakdowns for various block ciphers using CPU, NCP, and FlashVault (FV) with 256 KB input.}  
    \label{fig:block_cipher_performance_comparison}
    \vspace{-4mm}
\end{figure}

%\vspace{-1mm}
\subsection{Performance Evaluation}\label{sec:archi_evaluation_performance}
% \vspace{-1mm}

\noindent \textbf{1) Block Cipher:} Fig.~\ref{fig:block_cipher_performance_comparison} presents the latency of program and read operations for 256 KB input data.
Among the three systems, encryption on the host CPU incurs the highest latency due to data transfer and software overhead.
%When encryption is performed on the host CPU, data and software overhead leads to the highest latency among the three systems.
%When encryption is performed on the host CPU, the overhead from data transfer and software processing leads to the highest latency among the three systems.
The NCP architecture employs dedicated hardware near the SSD controller and removes software stack overhead, resulting in average latency improvements of 36.9\% and 43.2\% for program and read operations, respectively, over the real-machine system.
FlashVault integrates cryptographic engines within the NAND die and performs encryption directly without traversing the DRAM, achieving additional latency reductions of 28.8\% and 37.5\%, on average, for program (with encryption) and read (with decryption) operations, respectively, compared to NCP.
%, thereby achieving further average latency reductions of X\% and Y\% for program and read operations, respectively, compared to NCP.

\begin{figure}[t]
    \centering
    \includegraphics[scale=0.72]{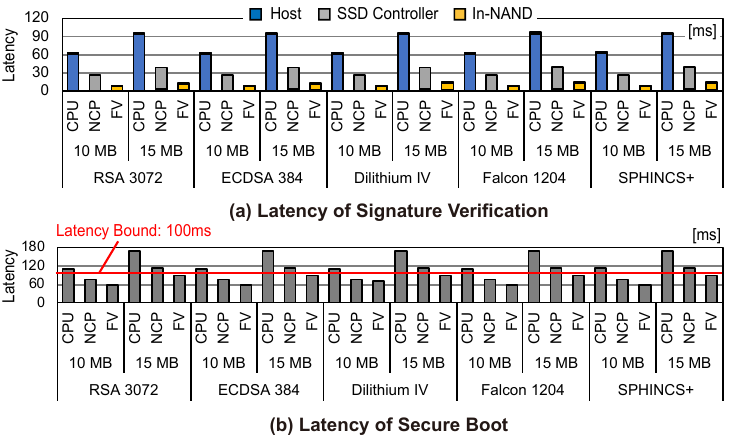}\vspace{-2mm} 
    \caption{Latencies of (a) signature verification and (b) end-to-end secure boot, evaluated with 10 MB and 15 MB firmware images.
    }  
    \label{fig:pkc_verification_performance}
    \vspace{-3mm}
\end{figure}

\noindent \textbf{2) Signature Verification:} Fig.~\ref{fig:pkc_verification_performance}-(a) presents the signature verification latency for 10~MB and 15~MB firmware images.
FlashVault achieves 1.9$\times$ and 1.3$\times$ lower latency, on average, compared to the real-machine system and the NCP architecture, respectively.
Fig.~\ref{fig:pkc_verification_performance}-(b) shows the total latency of secure boot, including boot code loading.
As discussed in Section~\ref{sec:moti_diverse_algorithms}, commercial SSDs are typically required to complete secure boot within 100 ms after power stabilization.
The host CPU–based verification fails to meet this constraint for all algorithms across the tested firmware sizes.
The NCP architecture meets the constraint for all algorithms at 10~MB but fails at 15~MB.
In contrast, FlashVault completes secure boot within 100~ms for all algorithms and image sizes, demonstrating its capability to securely boot even large firmware images used to support advanced features.

\begin{figure}[t]
    \centering
    \includegraphics[scale=0.70]{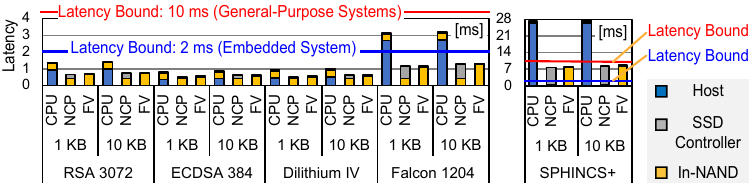}\vspace{-2mm} 
    \caption{Latency of signature generation evaluated with 1 KB and 10 KB inputs.
    Red and blue lines indicate latency bounds for anti-tamper logging in general-purpose and embedded systems, respectively.}  
    \label{fig:pkc_sining_performance}
    \vspace{-4mm}
\end{figure}

\noindent \textbf{3) Signature Generation:} Fig.~\ref{fig:pkc_sining_performance} shows the signature generation latency for 1~KB and 10~KB inputs.
%\samnotes{Seems it is Fig. 14?? But maybe not?? Fig. 14 is break down??}
FlashVault achieves 2.9$\times$ and 3.1$\times$ lower latency on average than the real-machine system for 1~KB and 10~KB, respectively. 
Anti-tamper logging, a representative application of signature generation, is typically required to complete within 2 to 15 ms in commercial SSDs, as discussed in Section~\ref{sec:sining_motivation}.
Specifically, real-time embedded systems typically require completion within 2 ms~\cite{emlog_logging}, while general-purpose systems tolerate up to 15 ms~\cite{ tamper_logging_size}.
However, the real-machine system exceeds the latency bounds of both systems when executing PQC algorithms.
In contrast, FlashVault meets the general-purpose latency bounds for all PQC algorithms and satisfies the stricter embedded constraint for all PQC algorithms except SPHINCS+.
This indicates that FlashVault supports most cryptographic algorithms in both embedded and general-purpose environments.
%As the evaluation used small input sizes, the DRAM access overhead in the NCP architecture is insignificant, resulting in a modest performance gain of approximately 0.8 to 1.0\% for FlashVault over NCP.
%The small input sizes used in the evaluation, which are representative of typical log entries, result in negligible DRAM access overhead in the NCP architecture. Consequently, FlashVault achieves a modest 0.8–1.0% performance gain over NCP. 
Since the evaluation uses small input sizes representative of typical log entries, the DRAM access overhead in the NCP is negligible, resulting in a modest performance gain of 0.2 to 1.0\% for FlashVault over NCP.

%\samnotes{So why not show larger input size results as well. Not only this, but overall, sensitivity analysis seems to be needed.}

%\noindent \textbf{4) Architectural Comparison:}
%\samnotes{I don't see a need for this part.}
%While NCP improves latency by placing cryptographic engines near the data source, it incurs additional area overhead near the SSD controller and increases the risk of exposing plaintext data and keys to DRAM.
%FlashVault, by contrast, integrates cryptographic engines into unused space within the 4D V-NAND chip, minimizing data exposure and eliminating extra area cost.

%NCP achieves faster performance than the real-machine system by performing cryptographic operations near the data source. 
%However,  it requires additional area near the SSD controller to accommodate cryptographic engines.
%Moreover, since encryption keys and plaintext data reside in DRAM prior to encryption, they are vulnerable to physical attacks discussed in Section~\ref{sec:threat_model}.
%In contrast, FlashVault integrates cryptographic engines into the unused space in the 4D V-NAND chip, delivering high performance without additional area cost. 
%By executing cryptographic operations directly within the NAND chip, FlashVault minimizes the exposure of plaintext data and keys to DRAM.

\begin{table}[t]\centering
\caption{Latency of cryptographic operations in a steady-state SSD}
% and its overhead measured relative to the case without GC/WL}
\vspace{-1mm}
\label{tab:gc_wl_overhead}
\scalebox{0.74}{
\begin{tabular}{cccc}
\hline\hline
\textbf{Scheme}                                                                   & \textbf{Operation}                                               & \textbf{\begin{tabular}[c]{@{}c@{}}Input \\ Size\end{tabular}} & \textbf{\begin{tabular}[c]{@{}c@{}}Cryptographic Algorithms\\ (Overall Latency {[}ms{]}; Overhead {[}\%{]})\end{tabular}}                                                            \\ 
\hline\hline
\multirow{2}{*}{\textbf{\begin{tabular}[c]{@{}c@{}}Block \\ Cipher\end{tabular}}} & \begin{tabular}[c]{@{}c@{}}Program\\ (w/ Encrypt)\end{tabular}   & 256 KB                                                         & \begin{tabular}[c]{@{}c@{}}3 DES (2.0; 31.7), HIGHT (1.7; 39.4), \\ IDEA (1.5; 46.6), Serpant (1.9; 35.3),  \\ SM4 (1.3; 55.9), Camellia (1.3; 55.6), AES (1.3; 58.0)\end{tabular} \\ 
\cline{2-4} 
                                                                                  & \begin{tabular}[c]{@{}c@{}}Read\\ (w/ Decrypt)\end{tabular}      & 256 KB                                                         & \begin{tabular}[c]{@{}c@{}}3 DES (1.6; 38.9), HIGHT (1.3; 53.2), \\ IDEA (1.1; 68.5), Serpant (1.5; 43.8), \\ SM4 (0.9; 92.9), Camellia (0.9; 90.1), AES (0.9; 96.2)\end{tabular}  \\ 
\hline\hline
\multirow{2}{*}{\textbf{\begin{tabular}[c]{@{}c@{}}PKC\\ \& PQC\end{tabular}}}    & \begin{tabular}[c]{@{}c@{}}Signature\\ Verification\end{tabular} & 15 MB                                                          & \begin{tabular}[c]{@{}c@{}}RSA 3072(13.5; 3.5), ECDSA 384 (13.5; 3.6), \\ Dilitium IV (13.6, 3.5),\\ Falcon 1024 (13.8; 3.5) , SPHINCS +  (13.7, 3.5)\end{tabular}                 \\ 
\cline{2-4} 
                                                                                  & \begin{tabular}[c]{@{}c@{}}Signature\\ Generation\end{tabular}   & 10 KB                                                          & \begin{tabular}[c]{@{}c@{}}RSA 3072(1.6; 116.7), ECDSA 384 (1.5; 154.0), \\ Dilitium IV (1.5; 155.3),\\ Falcon 1024 (2.2; 76.1) , SPHINCS +  (9.5; 12.1)\end{tabular}              \\ 
\hline\hline
\end{tabular}
}
\vspace{-3mm}
\end{table}

\noindent \textbf{4) FTL Overhead:} All prior evaluations were conducted under best-case conditions, with the system freshly initialized and no background FTL activities such as GC and WL.
In this state, the baseline FTL latency is approximately 4.6 $\mu$s for a 1 KB input and increases sub-linearly with larger input sizes. 
This trend is attributed to the proportional growth in processing time for address translation and data buffering. 
To assess the impact of GC and WL, we implemented a steady-state environment in SimpleSSD~\cite{modern_ssd_amber} by filling the SSD with dummy data and repeatedly overwriting random locations.
%We modified the \texttt{doGarbageCollection()} function in `page\_mapping.cc' of SimpleSSD to record timestamps at the start and end of each invocation.
Table~\ref{tab:gc_wl_overhead} presents latency and best-case-relative overhead under steady-state conditions, using the same inputs and workloads as in the previous evaluation.
For block ciphers, program operations incur an average FTL overhead of 46.1\%, while read operations show a higher 79.3\% overhead due to the relatively short NAND read latency, which makes the FTL more dominant in total latency.
For public-key operations, signature verification and signing show average overheads of 3.6\% and 127.6\%, respectively.
The verification incurs minimal overhead, as cryptographic latency dominates, whereas signing suffers from high overhead due to shorter execution time and smaller inputs.
SPHINCS+ deviates from the trend in signing overheads due to substantial cryptographic latency, which is caused by hash tree traversal and high computational complexity.
Although not shown in the table, verification with a 10 MB input and signing with a 1 KB input result in average GC/WL overheads of 5.6\% and 87.6\%, respectively.

%\section{FTL Overhead}\label{sec:ftl_overhead}

%\vspace{-1mm}
\section{Conclusion}\label{sec:conclusion}
%\samnotes{still working on this?}
%In this paper, we introduced FlashVault, 
We present FlashVault, a novel in-NAND self-encryption architecture that embeds reconfigurable cryptographic engines directly beneath the 4D NAND flash, enabling secure data processing with zero area overhead. 
By leveraging unused silicon area under vertically-stacked flash cell arrays, FlashVault supports diverse cryptographic algorithms, including block ciphers, public-key cryptography, and post-quantum cryptography, without relying on the host CPU. 
To enable this reconfigurability, we thoroughly broke down compute primitives in various cryptographic algorithms, so that the cryptographic engine in FlashVault can support all required primitives with reusable compute units.
FlashVault achieves 1.46$\sim$3.45$\times$ and 1.02$\sim$2.01$\times$ performance gains over host-based encryption and near-core processing architectures (i.e., with cryptographic engines placed near the SSD controller), respectively, across diverse cryptographic algorithms.

%\vspace{1mm}
%\section{Acknowledgments}
%This research was recognized with an award at the 31st Samsung HumanTech Award.
%\vspace{1mm}
%I sincerely thank my co-authors for their contributions to this work.
%I am especially grateful to Professors Sungjin Lee and Sam H. Noh for their invaluable advice and support. 
%It was a great honor to receive their thoughtful guidance.
%I am also deeply grateful to Professors Yeseong Kim and Jaeha Kung for their invaluable guidance and unwavering support.

% 노석환 박사, 아래 실험 결과를 기반으로 한 결론들 채워주면 될 듯 해요

%\section*{Acknowledgements}
%This document is an updated version of HPCA 2022 and 2023, which, in
%turn, has been derived from two previous conferences, in particular
%HPCA 2021 and MICRO 2021, which, in turn, are derived from past MICRO,
%HPCA, ISCA, and ASPLOS conferences.

%%%%%%% -- PAPER CONTENT ENDS -- %%%%%%%%

%%%%%%%%% -- BIB STYLE AND FILE -- %%%%%%%%
\bibliographystyle{IEEEtranS}
\bibliography{refs}

@misc{botan,
  author = {Botan},
  title = {{Botan: Crypto and TLS for Modern C++}},
  howpublished = "\url{https://botan.randombit.net/}",
  year = {2024}, 
 note = "[Online; accessed 02-June-2025]"
}

@misc{oecd_data_utilization,
  author       = {{OECD}},
  title        = {The Path to Becoming a Data-Driven Public Sector},
  howpublished = {\url{https://doi.org/10.1787/059814a7-en}},
  year         = {2019},
  note         = {Online; accessed 17-June-2025},
}

@misc{mckinsey_data_collect,
  author       = {McKinsey \& Company},
  title        = {The value of getting personalization right—or wrong—is multiplying},
  howpublished = {\url{https://www.mckinsey.com/business-functions/marketing-and-sales/our-insights/the-value-of-getting-personalization-right-or-wrong-is-multiplying}},
  year         = {2021},
  note         = {Online; accessed 17-June-2025}
}

@article{hft_damage,
  author  = {C. C. Moallemi and {\.I}. Sağlam},
  title   = {OR Forum—The Cost of Latency in High‑Frequency Trading},
  journal = {Operations Research},
  year    = {2013},
  volume  = {61},
  number  = {5},
  pages   = {1070--1086},
}

@article{blockchain_damage,
  author  = {Weizhao Tang and Lucianna Kiffer and Giulia Fanti and Ari Juels},
  title   = {Strategic Latency Reduction in Blockchain Peer-to-Peer Networks},
  journal = {Proceedings of the ACM on Measurement and Analysis of Computing Systems (POMACS)},
  volume  = {7},
  number  = {2},
  pages   = {1--33},
  month   = {Jun},
  year    = {2023},
  doi     = {10.1145/3589976},
}

@misc{segment_data_collect,
  author       = {Twilio Segment},
  title        = {The State of Personalization 2023},
  howpublished = {\url{https://gopages.segment.com/rs/667-MPQ-382/images/TS-CNT-Report-The%20State%20of%20Personalization%2023.pdf}},
  year         = {2023},
  note         = {Online; accessed 17-June-2025},
}

@misc{equifax_leak,
  author       = {Federal Trade Commission},
  title        = {Equifax Data Breach Settlement},
  howpublished = {\url{https://consumer.ftc.gov/consumer-alerts/2019/07/equifax-data-breach-settlement-what-you-should-know}},
  year         = {2019},
  note         = {Online; accessed 17-June-2025},
}

@misc{target_leak,
  author       = {HashedOut},
  title        = {Cost of 2013 Target Data Breach Nears \$3 Million},
  howpublished = {\url{https://www.thesslstore.com/blog/2013-target-data-breach-settled/},
  year         = {2017},
  note         = {Online; accessed 17-June-2025},
}}

@misc{dilithium,
  author = {Bai, Shi and Ducas, L\'eo and Kiltz, Eike and Lepoint, Tancr\`ede and 
          Lyubashevsky, Vadim and Schwabe, Peter and Seiler, Gregor and Stehl\'e, Damien},
  title = {{CRYSTALS-Dilithium Algorithm Specifications and Supporting Documentation}},
  howpublished = "\url{https://pq-crystals.org/dilithium/data/dilithium-specification-round3-20210208.pdf}",
  year = {2021}, 
 note = "[Online; accessed 02-July-2025]"
}

@misc{liboqs,
  author = {{Open Quantum Safe}},
  title = {{liboqs}},
  howpublished = "\url{https://openquantumsafe.org/liboqs/}",
  year = {2024}, 
 note = "[Online; accessed 02-July-2025]"
}

@misc{falcon,
  author = {{Open Quantum Safe}},
  title = {{liboqs}},
  howpublished = "\url{https://openquantumsafe.org/liboqs/}",
  year = {2024}, 
 note = "[Online; accessed 02-July-2025]"
}

@misc{falcon_first,
  author = {{Pierre-Alain Fouque and Jeffrey Hoffstein and Paul Kirchner and Vadim Lyubashevsky and Thomas Pornin and Thomas Prest and Thomas Ricosset and Gregor Seiler and William Whyte and Zhenfei Zhang}},
  title = {{Falcon: Fast-Fourier Lattice-based Compact Signatures over NTRU}},
  howpublished = "\url{https://api.semanticscholar.org/CorpusID:231637439}",
  year = {2019}, 
 note = "[Online; accessed 06-July-2025]"
}

@misc{openssl,
  author = {{OpenSSL}},
  title = {{The Open Source toolkit for SSL/TLS}},
  howpublished = "\url{https://www.openssl.org/}",
  year = {1998}, 
 note = "[Online; accessed 02-July-2025]"
}

@ARTICLE{power_supply,
  author={Zambelli, Cristian and Micheloni, Rino and Crippa, Luca and Zuolo, Lorenzo and Olivo, Piero},
  journal={IEEE Transactions on Device and Materials Reliability}, 
  title={Impact of the NAND Flash Power Supply on Solid State Drives Reliability and Performance}, 
  year={2018},
  volume={18},
  number={2},
  pages={247-255},
  doi={}}

@INPROCEEDINGS{power_dominant_program,
  author={Mohan, Vidyabhushan and Gurumurthi, Sudhanva and Stan, Mircea R.},
  booktitle={Design, Automation \& Test in Europe Conference \& Exhibition (DATE)}, 
  title={FlashPower: A detailed power model for NAND flash memory}, 
  year={2010},
  volume={},
  number={},
  pages={502-507},
  doi={}}

@INPROCEEDINGS{thermal_performance,
  author={Zambelli, Cristian and Crippa, Luca and Micheloni, Rino and Olivo, Piero},
  booktitle={International Integrated Reliability Workshop (IIRW)}, 
  title={Cross-Temperature Effects of Program and Read Operations in 2D and 3D NAND Flash Memories}, 
  year={2018},
  volume={},
  number={},
  pages={1-4},
  doi={2}}

@INPROCEEDINGS{cache_attack2,
  author={{Osvik, Dag Arne and Shamir, Adi and Tromer, Eran}},
  booktitle={IEEE Symposium on Security and Privacy (SP)}, 
  title={{Cache Attacks and Countermeasures: The Case of AES}}, 
  year={2006},
  volume={},
  number={},
  pages={200--213},
  doi={}}

@INPROCEEDINGS{sphincs,
  author={Bernstein, Daniel J. and H\"{u}lsing, Andreas and K\"{o}lbl, Stefan and Niederhagen, Ruben and Rijneveld, Joost and Schwabe, Peter},
  booktitle={ACM SIGSAC Conference on Computer and Communications Security (CCS)}, 
  title={{The SPHINCS+ Signature Framework}}, 
  year={2019},
  volume={},
  number={},
  pages={2129–2146},
  doi={10.1145/3319535.3363229}}

@misc{sha,
  author = {{National Institute of Standards and Technology (NIST)}},
  title = {{Secure Hash Standard (SHS)}},
  howpublished = "\url{https://files.floridados.gov/media/704729/fips-pub-180.pdf}",
  year = {2024}, 
  note = "[Online; accessed 06-June-2025]"
}

@misc{secure_boot_100ms,
  author = {{K. Nasim, E. Pang, J. Carreno, S. Kareti, K. C., R. Wahler, E. Marando, R. Rao, R. Kumar, and R. Schoepflin}},
  title = {{<Project Ersa AMC> Design Specification}},
  howpublished = "\url{https://www.opencompute.org/documents/project-ersa-amc-ocp-draft-v1-0-final-pdf-1}",
  year = {2024}, 
  note = "[Online; accessed 01-July-2025]"
}

@misc{texas_boot_size,
  author = {{Texas Instruments}},
  title = {{AM62P Boot Overview}},
  howpublished = "\url{https://e2e.ti.com/cfs-file/__key/communityserver-discussions-components-files/791/Sitara_5F00_AM62P_5F00_Boot_5F00_Overview.pdf}",
  year = {2023}, 
  note = "[Online; accessed 01-July-2025]"
}

@misc{SK_Hynix_ssd_product,
  author = {{SK Hynix}},
  title = {{SK hynix PE8111 E1.L NVMe TCG Opal SSC SED}},
  howpublished = "\url{https://csrc.nist.gov/CSRC/media/projects/cryptographic-module-validation-program/documents/security-policies/140sp4410.pdf}",
  year = {2022}, 
  note = "[Online; accessed 25-June-2025]"
}

@ARTICLE{ieee_std1667,
  author={{IEEE}},
  journal={{IEEE Std 1667-2015 (Revision of IEEE Std 1667-2009)}}, 
  title={{IEEE Standard for Discovery, Authentication, and Authorization in Host Attachments of Storage Devices}}, 
  year={2016},
  volume={},
  number={},
  pages={1-230},
  doi={10.1109/IEEESTD.2016.7389949}}

@misc{tcg_ssc,
  author = {{Trusted Computing Group}},
  title = {{TCG Storage Security Subsystem Class (SSC): Opal}},
  howpublished = "\url{https://trustedcomputinggroup.org/wp-content/uploads/TCG-Storage-Opal-SSC-v2.30_pub.pdf}",
  year = {2025}, 
  note = "[Online; accessed 20-June-2025]"
}

@misc{intel_aes,
  author = {{Intel}},
  title = {{What Type of Encryption Do Intel® SSDs Use?}},
  howpublished = "\url{https://www.intel.com/content/www/us/en/support/articles/000036098/memory-and-storage.html}",
  year = {2022}, 
  note = "[Online; accessed 20-June-2025]"
}

@INPROCEEDINGS{dilithium_paper,
  author={Ducas, L\'eo and Kiltz, Eike and Lepoint, Tancr\`ede and Lyubashevsky, Vadim and Schwabe, Peter and Seiler, Gregor and Stehl\'e, Damien},
  booktitle={IACR Transactions on Cryptographic Hardware and Embedded Systems (TCHES)}, 
  title={{CRYSTALS-Dilithium: A Lattice-Based Digital Signature Scheme}}, 
  year={2018},
  volume={},
  number={},
  pages={238-268},
  doi={doi.org/10.13154/tches.v2018.i1.238-2683}}

@misc{china_sm2,
  author = {{State Cryptography Administration}},
  title = {{ublic key cryptographic algorithm SM2 based on elliptic curves}},
  howpublished = "\url{https://www.cnnic.com.cn/ScientificResearch/LeadingEdge/soea/SM2/201312/t20131204_43349.htm}",
  year = {2013}, 
  note = "[Online; accessed 25-June-2025]"
}

@misc{ecc_bsi,
  author = {{Federal Office for Information Security (BSI)}},
  title = {{Elliptic Curve Cryptography}},
  howpublished = "\url{https://www.techpowerup.com/ssd-specs/sk-hynix-gold-p31-2-tb.d443}",
  year = {2018}, 
  note = "[Online; accessed 25-June-2025]"
}

@misc{kvkk,
  author = {{Kişisel Verileri Koruma Kanunu (KVKK)}},
  title = {{Personal Data Protection Authority}},
  howpublished = "\url{https://www.kvkk.gov.tr/SharedFolderServer/CMSFiles/1d7c2f99-be2c-4971-a304-0a1eb3586bd1.pdf}",
  year = {2024}, 
 note = "[Online; accessed 24-June-2025]"
}

@misc{gost,
  author = {{Vasily Dolmatov}},
  title = {{GOST R 34.12-2015}: Block Cipher "Kuznyechik"},
  howpublished = "\url{https://datatracker.ietf.org/doc/rfc7801/}",
  year = {2016}, 
 note = "[Online; accessed 24-June-2025]"
}

@misc{japan_camellia,
  author = {{Daniel Hounslow}},
  title = {{Japan-Data Protection Overview}},
  howpublished = "\url{https://www.dataguidance.com/notes/japan-data-protection-overview}",
  year = {2023}, 
 note = "[Online; accessed 24-June-2025]"
}

@misc{lpddr3,
  author = {{Micron}},
  title = {{Mobile LPDDR3 SDRAM: 178-Ball,
Single-Channel Mobile LPDDR3 SDRAM Features}},
  howpublished = "\url{www.micron.com/products/dram/lpdram/16Gb}",
  year = "2014"
}

@misc{nand_roadmap,
  author = {{Dick James and Jeongdong Choe}},
  title = {{NAND Flash Technology}},
  howpublished = "\url{https://www.techinsights.com/ko/node/30189}",
  year = {2019}, 
  note = "[Online; accessed 08-June-2025]"
}

@INPROCEEDINGS{14nm_nand,
  author={Lee, Seungjae and Lee, Jin-yub and Park, Il-han and Park, Jongyeol and Yun, Sung-won and Kim, Min-su and Lee, Jong-hoon and Kim, Minseok and Lee, Kangbin and Kim, Taeeun and Cho, Byungkyu and Cho, Dooho and Yun, Sangbum and Im, Jung-no and Yim, Hyejin and Kang, Kyung-hwa and Jeon, Suchang and Jo, Sungkyu and Ahn, Yang-lo and Joe, Sung-Min and Kim, Suyong and Woo, Deok-kyun and Park, Jiyoon and Park, Hyun-wook and Kim, Youngmin and Park, Jonghoon and Choi, Yongsu and Hirano, Makoto and Ihm, Jeong-Don and Jeong, Byunghoon and Lee, Seon-Kyoo and Kim, Moosung and Lee, Hokil and Seo, Sungwhan and Jeon, Hongsoo and Kim, Chan-ho and Kim, Hyunggon and Kim, Jintae and Yim, Yongsik and Kim, Hoosung and Byeon, Dae-Seok and Yang, Hyang-Ja and Park, Ki-Tae and Kyung, Kye-hyun and Choi, Jeong-Hyuk},
  booktitle={IEEE International Solid-State Circuits Conference (ISSCC)}, 
  title={{A 128Gb 2b/cell NAND flash memory in 14nm technology with tPROG=640µs and 800MB/s I/O rate}}, 
  year={2016},
  volume={},
  number={},
  pages={138-139},
  doi={10.1109/ISSCC.2016.7417945}}

@INPROCEEDINGS{modern_ssd_amber,
  author={Gouk, Donghyun and Kwon, Miryeong and Zhang, Jie and Koh, Sungjoon and Choi, Wonil and Kim, Nam Sung and Kandemir, Mahmut and Jung, Myoungsoo},
  booktitle={IEEE/ACM International Symposium on Microarchitecture (MICRO)}, 
  title={{Amber: Enabling Precise Full-System Simulation with Detailed Modeling of All SSD Resources}}, 
  year={2018},
  volume={},
  number={},
  pages={469-481},
  doi={10.1109/MICRO.2018.00045}}

@ARTICLE{samsung_3d_vnand,
  author={Park, Ki-Tae and Nam, Sangwan and Kim, Daehan and Kwak, Pansuk and Lee, Doosub and Choi, Yoon-He and Choi, Myung-Hoon and Kwak, Dong-Hun and Kim, Doo-Hyun and Kim, Min-Su and Park, Hyun-Wook and Shim, Sang-Won and Kang, Kyung-Min and Park, Sang-Won and Lee, Kangbin and Yoon, Hyun-Jun and Ko, Kuihan and Shim, Dong-Kyo and Ahn, Yang-Lo and Ryu, Jinho and Kim, Donghyun and Yun, Kyunghwa and Kwon, Joonsoo and Shin, Seunghoon and Byeon, Dae-Seok and Choi, Kihwan and Han, Jin-Man and Kyung, Kye-Hyun and Choi, Jeong-Hyuk and Kim, Kinam},
  journal={IEEE Journal of Solid-State Circuits (JSSC)}, 
  title={{Three-Dimensional 128 Gb MLC Vertical nand Flash Memory With 24-WL Stacked Layers and 50 MB/s High-Speed Programming}}, 
  year={2015},
  volume={50},
  number={1},
  pages={204-213},
  doi={10.1109/JSSC.2014.2352293}}

@ARTICLE{ecdsa,
  author={ohnson, Don and Menezes, Alfred and Vanstone, Scott},
  journal={{International Journal of Information Security (Int. J. Inf. Secur.)}}, 
  title={{The Elliptic Curve Digital Signature Algorithm (ECDSA)}}, 
  year={2001},
  volume={1},
  number={1},
  pages={36-63},
  doi={10.1007/s102070100002}}

@INPROCEEDINGS{Hynix_3d_vnand,
  author={Choi, Eun-Seok and Park, Sung-Kye},
  booktitle={International Electron Devices Meeting (IEDM)}, 
  title={{Device considerations for high density and highly reliable 3D NAND flash cell in near future}}, 
  year={2012},
  volume={},
  number={},
  pages={9.4.1-9.4.4},
  doi={10.1109/IEDM.2012.6479011}}

@article{rsa_first,
  author    = {Rivest, Ronald L. and Shamir, Adi and Adleman, Leonard},
  title     = {A Method for Obtaining Digital Signatures and Public-Key Cryptosystems},
  journal   = {Communications of the ACM (CACM)},
  volume    = {21},
  number    = {2},
  pages     = {120-126},
  year      = {1978},
  doi       = {10.1145/359340.359342}
}

@INPROCEEDINGS{idea,
  author={{Lai, Xuejia and Massey, James L.}},
  booktitle={Workshop on the theory and application of cryptographic techniques on Advances in cryptology (EUROCRYPT)}, 
  title={{A proposal for a new block encryption standard}}, 
  year={1991},
  volume={},
  number={},
  pages={389–404},
  doi={}}

@INPROCEEDINGS{keccak_permutation,
  author={{Ling Song and Guohong Liao and Jian Guo}},
  booktitle={Advances in Cryptology (CRYPTO)}, 
  title={{Non-full Sbox Linearization: Applications to Collision Attacks on Round-Reduced Keccak}}, 
  year={2017},
  volume={},
  number={},
  pages={428-451},
  doi={}}

@INPROCEEDINGS{serpant,
  author={{Biham, Eli and Anderson, Ross J. and Knudsen, Lars R.}},
  booktitle={International Workshop on Fast Software Encryption (FSE)}, 
  title={{Serpent: A New Block Cipher Proposal}}, 
  year={1998},
  volume={},
  number={},
  pages={222–238},
  doi={}}

@INPROCEEDINGS{micron_3d_nand,
  author={Tanaka, Tomoharu and Helm, Mark and Vali, Tommaso and Ghodsi, Ramin and Kawai, Koichi and Park, Jae-Kwan and Yamada, Shigekazu and Pan, Feng and Einaga, Yuichi and Ghalam, Ali and Tanzawa, Toru and Guo, Jason and Ichikawa, Takaaki and Yu, Erwin and Tamada, Satoru and Manabe, Tetsuji and Kishimoto, Jiro and Oikawa, Yoko and Takashima, Yasuhiro and Kuge, Hidehiko and Morooka, Midori and Mohammadzadeh, Ali and Kang, Jong and Tsai, Jeff and Sirizotti, Emanuele and Lee, Eric and Vu, Luyen and Liu, Yuxing and Choi, Hoon and Cheon, Kwonsu and Song, Daesik and Shin, Daniel and Yun, Jung Hee and Piccardi, Michele and Chan, Kim-Fung and Luthra, Yogesh and Srinivasan, Dheeraj and Deshmukh, Srinivasarao and Kavalipurapu, Kalyan and Nguyen, Dan and Gallo, Girolamo and Ramprasad, Sumant and Luo, Michelle and Tang, Qiang and Incarnati, Michele and Macerola, Agostino and Pilolli, Luigi and De Santis, Luca and Rossini, Massimo and Moschiano, Violante and Santin, Giovanni and Tronca, Bernardino and Lee, Hyunseok and Patel, Vipul and Pekny, Ted and Yip, Aaron and Prabhu, Naveen and Sule, Purval and Bemalkhedkar, Trupti and Upadhyayula, Kiranmayee and Jaramillo, Camila},
  booktitle={IEEE International Solid-State Circuits Conference (ISSCC)}, 
  title={{A 768Gb 3b/cell 3D-floating-gate NAND flash memory}}, 
  year={2016},
  volume={},
  number={},
  pages={142-144},
  doi={10.1109/ISSCC.2016.7417947}}

@misc{gdpr,
  author = {{European Parliament}},
  title = {{General Data Protection Regulation (GDPR)}},
  howpublished = "\url{https://eur-lex.europa.eu/legal-content/EN/TXT/PDF/?uri=CELEX:32016R0679}",
  year = {2016}, 
  note = "[Online; accessed 20-June-2025]"
}

@misc{3des,
  author = {{P. Karn and B. Kaliski}},
  title = {{The ESP Triple DES Transform}},
  howpublished = "\url{https://www.rfc-editor.org/rfc/rfc1851.html}",
  year = {1995}, 
  note = "[Online; accessed 02-July-2025]"
}

@misc{ccpa,
  author = {{State of California Department of Justice}},
  title = {{California Consumer Privacy Act (CCPA)}},
  howpublished = "\url{https://oag.ca.gov/privacy/ccpa}",
  year = {2024}, 
  note = "[Online; accessed 20-June-2025]"
}

@misc{korea_act,
  author = {{Korea Legislation Research Institute}},
  title = {{Personal Information Protection Act}},
  howpublished = "\url{https://elaw.klri.re.kr/eng_service/lawView.do?hseq=62389&lang=ENG}",
  year = {2023}, 
  note = "[Online; accessed 20-June-2025]"
}

@misc{kingston_standard,
  author = {{Kingston Technology}},
  title = {{What is SSD Encryption and How Does It Work?}},
  howpublished = "\url{https://www.kingston.com/en/blog/data-security/how-ssd-encryption-works}",
  year = {2021}, 
  note = "[Online; accessed 20-June-2025]"
}

@misc{samsung_firmware_size,
  author = {{Samsung}},
  title = {{Download – Tools \& Software}},
  howpublished = "\url{https://semiconductor.samsung.com/consumer-storage/support/tools/}",
  year = {2025}, 
  note = "[Online; accessed 06-July-2025]"
}

@INPROCEEDINGS{tamper_logging_size,
  author={{Ahmad, Adil and Lee, Sangho and Peinado, Marcus}},
  booktitle={IEEE Symposium on Security and Privacy (SP)}, 
  title={{HARDLOG: Practical Tamper-Proof System Auditing Using a Novel Audit Device}}, 
  year={2022},
  volume={},
  number={},
  pages={1791-1807},
  doi={10.1109/SP46214.2022.9833745}}

@article{survey_pkc,
  author    = {Koblitz, Neal and Menezes, Alfred J.},
  title     = {A Survey of Public-Key Cryptosystems},
  journal   = {SIAM Review},
  volume    = {46},
  number    = {4},
  pages     = {599–634},
  year      = {2004},
  doi       = {10.1137/S0036144503439190}
}

@INPROCEEDINGS{emlog_logging,
  author={{Shepherd, Carlton and Akram, Raja Naeem and Markantonakis, Konstantinos}},
  booktitle={Information Security Theory and Practice}, 
  title={{EmLog: Tamper-Resistant System Logging for Constrained Devices with TEEs}}, 
  year={2018},
  volume={},
  number={},
  pages={75-92},
  doi={}}

@ARTICLE{ieee_aes_std,
  author={{IEEE}},
  journal={{IEEE Std 1619-2018 (Revision of IEEE Std 1619-2007)}}, 
  title={{IEEE Standard for Cryptographic Protection of Data on Block-Oriented Storage Devices}}, 
  year={2019},
  volume={},
  number={},
  pages={1-41},
  doi={10.1109/IEEESTD.2019.8637988}}

@ARTICLE{sm4_report,
  author={{Haoran Gong and Tailiang Ju}},
  journal={{Scientific Reports
}}, 
  title={{Distributed power analysis attack on SM4 encryption chip}}, 
  year={2024},
  volume={14},
  number={1007},
  issn = {2045-2322},
  pages={},
  doi={10.1038/s41598-023-50220-2}}

@ARTICLE{scale_factor,
  author={Huang, Wei and Rajamani, Karthick and Stan, Mircea R. and Skadron, Kevin},
  journal={{IEEE Micro}}, 
  title={{Scaling with Design Constraints: Predicting the Future of Big Chips}}, 
  year={2011},
  volume={31},
  number={4},
  pages={16-29},
  doi={10.1109/MM.2011.42}}

@INPROCEEDINGS{rif,
  author={Chun, Myoungjun and Lee, Jaeyong and Kim, Myungsuk and Park, Jisung and Kim, Jihong},
  booktitle={IEEE International Symposium on High-Performance Computer Architecture (HPCA)}, 
  title={RiF: Improving Read Performance of Modern SSDs Using an On-Die Early-Retry Engine}, 
  year={2024},
  volume={},
  number={},
  pages={643-656},
  doi={}}

@misc{Arm_cortex_r8,
  author = {{ARM}},
  title = {{Cortex-R8}},
  howpublished = "\url{https://developer.arm.com/Processors/Cortex-R8}",
  year = {}, 
  note = "[Online; accessed 08-July-2025]"
}

@misc{japan_ppc,
  author = {{DataGuidance}},
  title = {{Japan - Data Protection Overview}},
  howpublished = "\url{https://www.dataguidance.com/notes/japan-data-protection-overview}",
  year = {2023}, 
  note = "[Online; accessed 20-June-2025]"
}

@misc{cmvp,
  author = {{National Institute of Standards and Technology (NIST)}},
  title = {{Cryptographic Module Validation Program (CMVP)}},
  howpublished = "\url{https://csrc.nist.gov/projects/cryptographic-module-validation-program}",
  year = {2025}, 
  note = "[Online; accessed 24-June-2025]"
}

@misc{fips_140_2,
  author = {{National Institute of Standards and Technology (NIST)} },
  title = {{Security Requirements for Cryptographic Modules}},
  howpublished = "\url{https://csrc.nist.gov/pubs/fips/140-2/upd2/final}",
  year = {2002}, 
  note = "[Online; accessed 24-June-2025]"
}

@misc{fips_140_3,
  author = {{National Institute of Standards and Technology (NIST)} },
  title = {{Security Requirements for Cryptographic Modules}},
  howpublished = "\url{https://csrc.nist.gov/pubs/fips/140-2/upd2/final}",
  year = {2019}, 
  note = "[Online; accessed 24-June-2025]"
}

@ARTICLE{hash_parallel_survey,
  author={Atighehchi, Kevin and Rolland, Robert},
  journal={IEEE Transactions on Computers (TC)}, 
  title={Optimization of Tree Modes for Parallel Hash Functions: A Case Study}, 
  year={2017},
  volume={66},
  number={9},
  pages={1585-1598},
  doi={10.1109/TC.2017.2693185}}

@misc{kcmvp_example,
  author = {{Korea Internet and Security Agency (KISA)}},
  title = {{Korean Cryptographic Module Validation Program}},
  howpublished = "\url{https://seed.kisa.or.kr/kisa/kcmvp/EgovVerification.do}",
  year = {2023}, 
  note = "[Online; accessed 24-June-2025]"
}

@misc{samsung_secure_firmware,
  author = {{Samsung}},
  title = {{Samsung NVMe TCG Opal SSC SEDs PM1733/PM1735 Series}},
  howpublished = "\url{https://csrc.nist.gov/CSRC/media/projects/cryptographic-module-validation-program/documents/security-policies/140sp4238.pdf}",
  year = {2021}, 
  note = "[Online; accessed 20-June-2025]"
}

@misc{micron_secure_firmware,
  author = {{Micron}},
  title = {{MICRON 1100 SSD FIPS 140-2 Cryptographic Module Non-Proprietary Security Policy}},
  howpublished = "\url{https://csrc.nist.gov/csrc/media/projects/cryptographic-module-validation-program/documents/security-policies/140sp2848.pdf}",
  year = {2024}, 
  note = "[Online; accessed 20-June-2025]"
}

@misc{nist_pqc_altorithms,
  author = {{{National Institute of Standards and Technology (NIST)}}},
  title = {{NIST Releases First 3 Finalized Post-Quantum Encryption Standards}},
  howpublished = "\url{https://www.nist.gov/news-events/news/2024/08/nist-releases-first-3-finalized-post-quantum-encryption-standards}",
  year = {2024}, 
  note = "[Online; accessed 20-June-2025]"
}

@misc{ibm_tls,
  author = {{Hartmut Lonzer, Corne Lottering, and Vasfi Gucer}},
  title = {{IBM Storage FlashSystem 5200 Product Guide for IBM Storage Virtualize 8.6}},
  howpublished = "\url{https://www.redbooks.ibm.com/redpapers/pdfs/redp5617.pdf}",
  year = {2023}, 
  note = "[Online; accessed 20-June-2025]"
}

@misc{snia_cpu,
  author = {{Storage Networking Industry Association (SNIA)}},
  title = {{Storage Security: Encryption and Key Management}},
  howpublished = "\url{https://www.snia.org/sites/default/files/technical-work/whitepapers/SNIA-Encryption-Key-Management-Tech-Whitepaper.pdf}",
  year = {2023}, 
  note = "[Online; accessed 21-June-2025]"
}

@ARTICLE{cell_interference,
  author={Jae-Duk Lee and Sung-Hoi Hur and Jung-Dal Choi},
  journal={IEEE Electron Device Letters (EDL)}, 
  title={{Effects of floating-gate interference on NAND flash memory cell operation}}, 
  year={2002},
  volume={23},
  number={},
  pages={264-266},
  doi={10.1109/55.998871}}

@ARTICLE{pqc_small_length,
  author={Nejatollahi, Hamid and Dutt, Nikil and Ray, Sandip and Regazzoni, Francesco and Banerjee, Indranil and Cammarota, Rosario},
  journal={ACM Computing Surveys (CSUR)}, 
  title={{Post-Quantum Lattice-Based Cryptography Implementations: A Survey}}, 
  year={2019},
  volume={51},
  number={6},
  pages={1-41},
  doi={}}

@ARTICLE{modern_ssd_survey,
  author={Cai, Yu and Ghose, Saugata and Haratsch, Erich F. and Luo, Yixin and Mutlu, Onur},
  journal={Proceedings of the IEEE}, 
  title={{Error Characterization, Mitigation, and Recovery in Flash-Memory-Based Solid-State Drives}}, 
  year={2017},
  volume={105},
  number={},
  pages={1666-1704},
  doi={10.1109/JPROC.2017.2713127}}

@inproceedings{pif,
author = {Chun, Myungjun and Lee, Jaeyong and Lee, Sanggu and Kim, Myungsuk and Kim, Jihong},
title = {PiF: in-flash acceleration for data-intensive applications},
year = {2022},
isbn = {},
publisher = {},
address = {},
url = {},
doi = {10.1145/3538643.3539742},
booktitle = {ACM Workshop on Hot Topics in Storage and File Systems (HotStorage)},
pages = {106–112},
numpages = {7},
location = {},
series = {}
}

@INPROCEEDINGS{CUA_patent,
  author={Chia-Lin Hsiung and Yanbin An and Alexander Chu and Fumiaki Toyama},
  booktitle={US9922716B2 (US Patent)}, 
  title={{Architecture for CMOS under array}}, 
  year={2018},
  volume={},
  number={},
  pages={},
  doi={}}

@INPROCEEDINGS{camellia,
  author={{Aoki, Kazumaro and Ichikawa, Tetsuya and Kanda, Masayuki and Matsui, Mitsuru and Moriai, Shiho and Nakajima, Junko and Tokita, Toshio}},
  booktitle={International Workshop on Selected Areas in Cryptography (SAC)}, 
  title={{Camellia: A 128-Bit Block Cipher Suitable for Multiple Platforms - Design and Analysis}}, 
  year={2000},
  volume={},
  number={},
  pages={39–56},
  doi={}}

@INPROCEEDINGS{quantum_threat,
  author={Shor, P.W.},
  booktitle={Annual IEEE Symposium on Foundations of Computer Science (FOCS)}, 
  title={Algorithms for quantum computation: discrete logarithms and factoring}, 
  year={1994},
  volume={},
  number={},
  pages={124-134},
  doi={10.1109/SFCS.1994.365700}}

@INPROCEEDINGS{HIGHT,
  author={Hong, Deukjo and Sung, Jaechul and Hong, Seokhie and Lim, Jongin and Lee, Sangjin and Koo, Bon-Seok and Lee, Changhoon and Chang, Donghoon and Lee, Jesang and Jeong, Kitae and Kim, Hyun and Kim, Jongsung and Chee, Seongtaek},
  booktitle={International Conference on Cryptographic Hardware and Embedded Systems (CHES)}, 
  title={{HIGHT}: a new block cipher suitable for low-resource device}, 
  year={2006},
  volume={},
  number={},
  pages={46–59},
  doi={10.1007/11894063_4}}

@INPROCEEDINGS{anti_bus_probing2,
  author={Xu, Zhenyu and Mauldin, Thomas and Yao, Zheyi and Pei, Shuyi and Wei, Tao and Yang, Qing},
  booktitle={ACM/IEEE International Symposium on Computer Architecture (ISCA)}, 
  title={{A Bus Authentication and Anti-Probing Architecture Extending Hardware Trusted Computing Base Off CPU Chips and Beyond}}, 
  year={2020},
  volume={},
  number={},
  pages={749-761},
  doi={10.1109/ISCA45697.2020.00067}}

@inproceedings{wear_leveling_attack1,
author = {Wei, Michael and Grupp, Laura M. and Spada, Frederick E. and Swanson, Steven},
title = {{Reliably erasing data from flash-based solid state drives}},
year = {2011},
isbn = {},
publisher = {},
address = {},
booktitle = {USENIX Conference on File and Stroage Technologies (FAST)},
pages = {8},
numpages = {1},
location = {},
series = {}
}

@INPROCEEDINGS{power_loss_date,
  author={Ahmadian, Saba and Taheri, Farhad and Lotfi, Mehrshad and Karimi, Maryam and Asadi, Hossein},
  booktitle={Design, Automation \& Test in Europe Conference \& Exhibition (DATE)}, 
  title={Investigating power outage effects on reliability of solid-state drives}, 
  year={2018},
  volume={},
  number={},
  pages={207-212},
  doi={}}

@INPROCEEDINGS{power_loss_fast,
  author={Zheng, Mai and Tucek, Joseph and Qin, Feng and Lillibridge, Mark},
  booktitle={USENIX Conference on File and Storage Technologies (FAST)}, 
  title={Understanding the robustness of {SSDs} under power fault}, 
  year={2013},
  volume={},
  number={},
  pages={271–284},
  doi={}}

@article{power_attack,
  author = {Mark Randolph and William Diehl},
  title = {Power Side‑Channel Attack Analysis: A Review of 20 Years of Study for the Layman},
  journal = {Cryptography},
  year = {2020},
  volume = {4},
  number = {2},
}

@ARTICLE{power_attack2,
  author={Ramezanpour, Keyvan and Ampadu, Paul and Diehl, William},
  journal={IEEE Transactions on Computers (TC)}, 
  title={SCAUL: Power Side-Channel Analysis With Unsupervised Learning}, 
  year={2020},
  volume={69},
  number={11},
  pages={1626-1638},
  doi={}}

@inproceedings{em_attack,
  author       = {Jake Longo Galea and Elke De Mulder and Daniel Page and Michael Tunstall},
  title        = {SoC It to EM: ElectroMagnetic Side‑Channel Attacks on a Complex System‑on‑Chip},
  booktitle    = {Cryptographic Hardware and Embedded Systems (CHES)},
  series       = {Lecture Notes in Computer Science},
  volume       = {9293},
  pages        = {620--640},
  year         = {2015},
  publisher    = {},
  doi          = {},
  url          = {}
}

@misc{hynix_die_area,
  author = {{SK Hynix}},
  title = {{SK hynix 512Gb 176L TLC 3D NAND Internal Waveform Overview}},
  howpublished = "\url{https://library.techinsights.com/analysis-view/IWO-2206-801}",
  year = {2024}, 
  note = "[Online; accessed 19-June-2025]"
}

@article{em_attack2,
  author       = {Asanka Sayakkara and Nhien\-An Le-Khac and Mark Scanlon},
  title        = {A Survey of Electromagnetic Side‑Channel Attacks and Discussion on Their Case‑Progressing Potential for Digital Forensics},
  journal      = {Digital Investigation},
  volume       = {29},
  pages        = {43--54},
  year         = {2019},
  doi          = {},
  url          = {}
}

@INPROCEEDINGS{firmware_attack,
  author={Meijer, Carlo and van Gastel, Bernard},
  booktitle={IEEE Symposium on Security and Privacy (SP)}, 
  title={{Self-Encrypting Deception: Weaknesses in the Encryption of Solid State Drives}}, 
  year={2019},
  volume={},
  number={},
  pages={72-87},
  doi={10.1109/SP.2019.00088}}

@INPROCEEDINGS{Hynix_4d_nand,
  author={Park, Jae-Woo and Kim, Doogon and Ok, Sunghwa and Park, Jaebeom and Kwon, Taeheui and Lee, Hyunsoo and Lim, Sungmook and Jung, Sun-Young and Choi, Hyeongjin and Kang, Taikyu and Park, Gwan and Yang, Chul-Woo and Choi, Jeong-Gil and Ko, Gwihan and Shin, Jaehyeon and Yang, Ingon and Nam, Junghoon and Sohn, Hyeokchan and Hong, Seok-In and Jeong, Yohan and Choi, Sung-Wook and Choi, Changwoon and Shin, Hyun-Soo and Lim, Junyoun and Youn, Dongkyu and Nam, Sanghyuk and Lee, Juyeab and Ahn, Myungkyu and Lee, Hoseok and Lee, Seungpil and Park, Jongmin and Gwon, Kichang and Jeong, Woopyo and Choi, Jungdal and Kim, Jinkook and Jin, Kyo-Won},
  booktitle={IEEE International Solid-State Circuits Conference (ISSCC)}, 
  title={A 176-Stacked 512Gb 3b/Cell 3D-NAND Flash with 10.8Gb/mm2 Density with a Peripheral Circuit Under Cell Array Architecture}, 
  year={2021},
  volume={64},
  number={},
  pages={422-423},
  doi={}}

@misc{3d_4d_comparison_nand,
  author = {{Billy Tallis}},
  title = {2021 NAND Flash Updates from ISSCC: The Learning Towers of TLC and QLC},
  howpublished = "\url{https://www.anandtech.com/show/16491/flash-memory-at-isscc-2021}",
  year = {2021}, 
  note = "[Online; accessed 19-June-2025]"
}

@misc{4d_vnand_data_processing,
  author       = {Sung, Dukhyun and Lee, Donghyun and Ryu, Jaeheon and Lee, Jaejin},
  title        = {Computation-in-memory in three-dimensional memory device},
  howpublished = {U.S. Patent 11,461,266 B2},
  year         = {2022},
  note         = {}
}

@misc{4d_vnand_neuron_circuit,
  author       = {Fu-Chang Hsu and San Jose},
  title        = {{3D} cell and array structures},
  howpublished = {U.S. Patent Application US2024/0404598 A1},
  year         = {2024},
  note         = {}
}

@misc{tcg_sed,
  author = {{Trusted Computing Group}},
  title = {ARCHITECT’S GUIDE: DATA SECURITY USING TCG SELF-ENCRYPTING DRIVE TECHNOLOGY},
  howpublished = "\url{https://trustedcomputinggroup.org/wp-content/uploads/Architects-Guide-Data-Security-Using-TCG-Self-Encrypting-Drive-Technology.pdf}",
  year = {2013}, 
  note = "[Online; accessed 22-July-2024]"
}

@misc{rc_synopsys,
  author = {{Synopsys}},
  title = {Synopsys StarRC: Golden Signoff Parasitic Extraction},
  howpublished = "\url{https://news.synopsys.com/home?item=123041}",
  year = "2025",
  note = "[Online; accessed 16-July-2025]"
}

@ARTICLE{wallace_multiplier,
  author={Waters, Ron S. and Swartzlander, Earl E.},
  journal={IEEE Transactions on Computers (TC)}, 
  title={A Reduced Complexity Wallace Multiplier Reduction}, 
  year={2010},
  volume={59},
  number={8},
  pages={1134-1137},
  doi={10.1109/TC.2010.103}}

@ARTICLE{barrett_reduction,
  author={Knezevic, Miroslav and Vercauteren, Frederik and Verbauwhede, Ingrid},
  journal={IEEE Transactions on Computers (TC)}, 
  title={Faster Interleaved Modular Multiplication Based on Barrett and Montgomery Reduction Methods}, 
  year={2010},
  volume={59},
  number={12},
  pages={1715-1721},
  doi={10.1109/TC.2010.93}}

@misc{tech_insight_survey,
  author = {Tech Insights},
  title = {{Comparison: Latest 3D NAND Products from YMTC, Samsung, SK hynix and Micron}},
  howpublished = "\url{https://www.techinsights.com/ko/node/49939}",
  year = {2023}, 
  note = "[Online; accessed 23-June-2025]"
}

@misc{synopsys_icc,
  author = {Synopsys},
  title = {{IC Compiler II Implementation User Guide: Version L-2016.03}},
  year = {2025}, 
  note = "[Online; accessed 19-Feburary-2025]"
}

@misc{synopsys_pt,
  author = {{Synopsys}},
  title = {Synopsys PrimeTime PX Power Analysis Solution Achieves Broad Market Adoption},
  howpublished = "\url{https://news.synopsys.com/home?item=123041}",
  year = "2025",
  note = "[Online; accessed 16-Feburary-2025]"
}

@INPROCEEDINGS{decoupled_ssd,
  author={Kim, Jiho and Jung, Myoungsoo and Kim, John},
  booktitle={ACM/IEEE International Symposium on Computer Architecture (ISCA)}, 
  title={Decoupled SSD: Rethinking SSD Architecture through Network-based Flash Controllers}, 
  year={2023},
  volume={},
  pages={},
  numpages = {13},
  doi={}}

@INPROCEEDINGS{d_shield,
  author={Chowdhuryy, Md Hafizul Islam and Jung, Myoungsoo and Yao, Fan and Awad, Amro},
  booktitle={IEEE International Symposium on High-Performance Computer Architecture (HPCA)}, 
  title={D-Shield: Enabling Processor-side Encryption and Integrity Verification for Secure NVMe Drives}, 
  year={2023},
  volume={},
  number={},
  pages={908-921},
  doi={}}

@INPROCEEDINGS{hardi_ssd_isca,
  author    = {Rohan Mahapatra and Harsha Santhanam and Christopher Priebe and Hanyang Xu and Hadi Esmaeilzadeh},
  title     = {In-Storage Acceleration of Retrieval Augmented Generation as a Service},
  booktitle = {ACM/IEEE International Symposium on Computer Architecture (ISCA)},
  year      = {2025},
  pages     = {450--466},
  doi       = {10.1145/3695053.3731032},
  publisher = {},
  address   = {}
}

@misc{nist_hash,
  author = {Quynh Dang},
  title = {Recommendation for Applications Using Approved Hash Algorithms},
  howpublished = "\url{https://doi.org/10.6028/NIST.SP.800-107r1}",
  year = {2009}, 
  note = "[Online; accessed 16-July-2025]"
}

@misc{samng_current,
  author = {Samsung},
  title = {The future of NAND technology},
  howpublished = "\url{https://semiconductor.samsung.com/news-events/tech-blog/the-future-of-nand-technology/}",
  year = {2019}, 
  note = "[Online; accessed 14-June-2025]"
}

@misc{power_efficiency_improve,
  author = {Jaihyuk Song},
  title = {skip to contentMenu openSearch openNation choice page link
[Editorial] Extraordinary Innovation for a More Unforgettable World: The Story Behind Samsung’s Pioneering V-NAND Memory Solution},
  howpublished = "\url{https://news.samsung.com/global/editorial-extraordinary-innovation-for-a-more-unforgettable-world-the-story-behind-samsungs-pioneering-v-nand-memory-solution}",
  year = {2021}, 
  note = "[Online; accessed 16-June-2025]"
}

@INPROCEEDINGS{flagger,
  author={Pan, Xiurui and An, Yuda and Liang, Shengwen and Mao, Bo and Zhang, Mingzhe and Li, Qiao and Jung, Myoungsoo and Zhang, Jie},
  booktitle={ACM/IEEE International Symposium on Computer Architecture (ISCA)}, 
  title={Flagger: Cooperative Acceleration for Large-Scale Cross-Silo Federated Learning Aggregation}, 
  year={2024},
  volume={},
  number={},
  pages={915-930},
  doi={10.1109/ISCA59077.2024.00071}}

@inproceedings{aes_shift,
author = {Chen, Xiaoqi},
title = {{Implementing AES Encryption on Programmable Switches via Scrambled Lookup Tables}},
year = {2020},
isbn = {},
publisher = {},
address = {},
url = {},
doi = {10.1145/3405669.3405819},
booktitle = {ACM SIGCOMM Workshop on Secure Programmable Network Infrastructure (SPIN)},
pages = {8–14},
numpages = {},
location = {},
series = {}
}

@inproceedings{modern_ssd_asplos,
author = {{Park, Jisung and Kim, Myungsuk and Chun, Myoungjun and Orosa, Lois and Kim, Jihong and Mutlu, Onur}},
title = {{Reducing Solid-State Drive Read Latency by Optimizing Read-Retry}},
year = {2021},
isbn = {},
publisher = {},
address = {},
url = {},
doi = {10.1145/3445814.3446719},
booktitle = {ACM International Conference on Architectural Support for Programming Languages and Operating Systems (ASPLOS)},
pages = {702–716},
numpages = {15},
location = {},
series = {}
}

@article{ro_puf,
author = {Maiti, Abhranil and Schaumont, Patrick},
title = {Improved Ring Oscillator PUF: An FPGA-friendly Secure Primitive},
year = {2011},
issue_date = {},
publisher = {},
address = {},
volume = {24},
number = {2},
issn = {},
url = {},
doi = {10.1007/s00145-010-9088-4},
journal = {Journal of Cryptology},
month = {},
pages = {375–397},
numpages = {23}
}

@ARTICLE{ldpc,
  author={{Cui, Hangxuan and Lin, Jun and Wang, Zhongfeng}},
  journal={IEEE Transactions on Circuits and Systems I: Regular Papers (TCAS-I)}, 
  title={{An Improved Gradient Descent Bit-Flipping Decoder for LDPC Codes}}, 
  year={2019},
  volume={66},
  number={8},
  pages={3188-3200},
  doi={10.1109/TCSI.2019.2909653}}

@misc{dell_sed,
  author       = {Dell EMC},
  title        = {Self-Encrypting Drives in Dell EMC PowerEdge Servers with VMware vSphere},
  year         = {2020},
  howpublished = {\url{https://dl.dell.com/manuals/common/vmware_esxi_6x_7x_whitepaper7_en_us.pdf}},
  note         = {Accessed: July 7, 2025}
}

@misc{qnap_sed,
  author       = {QNAP Systems},
  title        = {How to Use Self-Encrypting Drives (SEDs) on Your QNAP NAS},
  year         = {2021},
  howpublished = {\url{https://www.qnap.com/en/how-to/tutorial/article/how-to-use-self-encrypting-drives-seds-on-your-qnap-nas}},
  note         = {Accessed: July 7, 2025}
}

@ARTICLE{storage_energy_latency,
  author={Kumar, Akshay and Tandon, Ravi and Clancy, T. Charles},
  journal={IEEE Transactions on Cloud Computing (TCC)}, 
  title={On the Latency and Energy Efficiency of Distributed Storage Systems}, 
  year={2017},
  volume={5},
  number={2},
  pages={221-233},
  doi={10.1109/TCC.2015.2459711}}

@inproceedings{ecssd,
author = {Li, Siqi and Tu, Fengbin and Liu, Liu and Lin, Jilan and Wang, Zheng and Kang, Yangwook and Ding, Yufei and Xie, Yuan},
title = {{ECSSD: Hardware/Data Layout Co-Designed In-Storage-Computing Architecture for Extreme Classification}},
year = {2023},
doi = {10.1145/3579371.3589093},
booktitle = {ACM/IEEE Annual International Symposium on Computer Architecture (ISCA)},
numpages = {14}
}

@Article{4d_nand_survey,
AUTHOR = {Goda, Akira},
TITLE = {{Recent Progress on 3D NAND Flash Technologies}},
JOURNAL = {Electronics},
VOLUME = {10},
YEAR = {2021},
NUMBER = {24},
ARTICLE-NUMBER = {3156},
DOI = {10.3390/electronics10243156}
}

@INPROCEEDINGS{aes_first,
  author={{Sanchez-Avila, C. and Sanchez-Reillol, R.}},
  booktitle={IEEE International Carnahan Conference on Security Technology (ICCST)}, 
  title={{The Rijndael block cipher (AES proposal) : a comparison with DES}}, 
  year={2001},
  volume={},
  number={},
  pages={229-234},
  doi={10.1109/CCST.2001.962837}}

@INPROCEEDINGS{samsung_7,
  author={Cho, Jiho and Kang, D. Chris and Park, Jongyeol and Nam, Sang-Wan and Song, Jung-Ho and Jung, Bong-Kil and Lyu, Jaedoeg and Lee, Hogil and Kim, Won-Tae and Jeon, Hongsoo and Kim, Sunghoon and Kim, In-Mo and Son, Jae-Ick and Kang, Kyoungtae and Shim, Sang-Won and Park, JongChul and Lee, Eungsuk and Kang, Kyung-Min and Park, Sang-Won and Lee, Jaeyun and Moon, Seung Hyun and Kwak, Pansuk and Jeong, ByungHoon and Lee, Cheon An and Kim, Kisung and Ko, Junyoung and Kwon, Tae-Hong and Lee, Junha and Lee, Yohan and Kim, Chaehoon and Lee, Myeong-Woo and Yun, Jeong-yun and Lee, HoJun and Choi, Yonghyuk and Hong, Sanggi and Park, JongHoon and Shin, Yoonsung and Kim, Hojoon and Kim, Hansol and Yoon, Chiweon and Byeon, Dae Seok and Lee, Seungjae and Lee, Jin-Yub and Song, Jaihyuk},
  booktitle={IEEE International Solid-State Circuits Conference (ISSCC)}, 
  title={{30.3 A 512Gb 3b/Cell 7th -Generation 3D-NAND Flash Memory with 184MB/s Write Throughput and 2.0Gb/s Interface}}, 
  year={2021},
  volume={64},
  number={},
  pages={426-428},
  doi={10.1109/ISSCC42613.2021.9366054}}

@inproceedings{evanesco,
author = {Kim, Myungsuk and Park, Jisung and Cho, Genhee and Kim, Yoona and Orosa, Lois and Mutlu, Onur and Kim, Jihong},
title = {{Evanesco: Architectural Support for Efficient Data Sanitization in Modern Flash-Based Storage Systems}},
year = {2020},
doi = {10.1145/3373376.3378490},
booktitle = {ACM Architectural Support for Programming Languages and Operating Systems (ASPLOS)},
pages = {1311–1326}
}

@inproceedings{3d_nand_reliability_nand,
author = {Shim, Youngseop and Kim, Myungsuk and Chun, Myoungjun and Park, Jisung and Kim, Yoona and Kim, Jihong},
title = {{Exploiting Process Similarity of 3D Flash Memory for High Performance SSDs}},
year = {2019},
doi = {10.1145/3352460.3358311},
booktitle = {IEEE/ACM International Symposium on Microarchitecture (MICRO)},
pages = {211–223}
}

@article{nand_pomacs,
author = {Luo, Yixin and Ghose, Saugata and Cai, Yu and Haratsch, Erich F. and Mutlu, Onur},
title = {{Improving 3D NAND Flash Memory Lifetime by Tolerating Early Retention Loss and Process Variation}},
year = {2018},
issue_date = {},
publisher = {},
address = {},
volume = {2},
number = {3},
url = {},
doi = {10.1145/3224432},
journal = { ACM on Measurement and Analysis of Computing Systems (POMACS)},
articleno = {37},
numpages = {48}
}

@INPROCEEDINGS{heatwatch,
  author={Luo, Yixin and Ghose, Saugata and Cai, Yu and Haratsch, Erich F. and Mutlu, Onur},
  booktitle={IEEE International Symposium on High Performance Computer Architecture (HPCA)}, 
  title={{HeatWatch: Improving 3D NAND Flash Memory Device Reliability by Exploiting Self-Recovery and Temperature Awareness}}, 
  year={2018},
  volume={},
  number={},
  pages={504-517},
  doi={10.1109/HPCA.2018.00050}}

@INPROCEEDINGS{samsung_6,
  author={Kang, Dongku and Kim, Minsu and Jeon, Su Chang and Jung, Wontaeck and Park, Jooyong and Choo, Gyosoo and Shim, Dong-kyo and Kavala, Anil and Kim, Seung-Bum and Kang, Kyung-Min and Lee, Jiyoung and Ko, Kuihan and Park, Hyun-Wook and Min, Byung-Jun and Yu, Changyeon and Yun, Sewon and Kim, Nahyun and Jung, Yeonwook and Seo, Sungwhan and Kim, Sunghoon and Lee, Moo Kyung and Park, Joo-Yong and Kim, James C. and Cha, Young San and Kim, Kwangwon and Jo, Youngmin and Kim, Hyunjin and Choi, Youngdon and Byun, Jindo and Park, Ji-hyun and Kim, Kiwon and Kwon, Tae-Hong and Min, Youngsun and Yoon, Chiweon and Kim, Youngcho and Kwak, Dong-Hun and Lee, Eungsuk and Hahn, Wook-ghee and Kim, Ki-sung and Kim, Kyungmin and Yoon, Euisang and Kim, Won-Tae and Lee, Inryoul and Moon, Seung hyun and Ihm, Jeongdon and Byeon, Dae Seok and Song, Ki-Whan and Hwang, Sangjoon and Kyung, Kye Hyun},
  booktitle={IEEE International Solid-State Circuits Conference (ISSCC)}, 
  title={{13.4 A 512Gb 3-bit/Cell 3D 6th-Generation V-NAND Flash Memory with 82MB/s Write Throughput and 1.2Gb/s Interface}}, 
  year={2019},
  volume={},
  number={},
  pages={216-218},
  doi={10.1109/ISSCC.2019.8662493}}

@INPROCEEDINGS{optimstore,
  author={Kim, Junkyum and Kang, Myeonggu and Han, Yunki and Kim, Yang-Gon and Kim, Lee-Sup},
  booktitle={IEEE International Symposium on High-Performance Computer Architecture (HPCA)}, 
  title={{OptimStore: In-Storage Optimization of Large Scale DNNs with On-Die Processing}}, 
  year={2023},
  volume={},
  number={},
  pages={611-623},
  doi={10.1109/HPCA56546.2023.10071024}}

@inproceedings {cache_attack1,
author = {Yarom, Yuval and Falkner, Katrina E.},
title = {Flush+Reload: A High Resolution, Low Noise, L3 Cache Side-Channel Attack},
booktitle = {USENIX Security Symposium (USENIX Security)},
year = {2014},
isbn = {978-1-931971-15-7},
pages = {719--732},
url = {},
publisher = {},
}

@INPROCEEDINGS{modern_ssd_isca,
  author={Tavakkol, Arash and Sadrosadati, Mohammad and Ghose, Saugata and Kim, Jeremie and Luo, Yixin and Wang, Yaohua and Mansouri Ghiasi, Nika and Orosa, Lois and Gómez-Luna, Juan and Mutlu, Onur},
  booktitle={ACM/IEEE International Symposium on Computer Architecture (ISCA)}, 
  title={{FLIN: Enabling Fairness and Enhancing Performance in Modern NVMe Solid State Drives}}, 
  year={2018},
  volume={},
  number={},
  pages={397-410},
  doi={10.1109/ISCA.2018.00041}}

@INPROCEEDINGS{anti_bus_probing,
  author={Xu, Zhenyu and Mauldin, Thomas and Yao, Zheyi and Pei, Shuyi and Wei, Tao and Yang, Qing},
  booktitle={ACM/IEEE International Symposium on Computer Architecture (ISCA)}, 
  title={{A Bus Authentication and Anti-Probing Architecture Extending Hardware Trusted Computing Base Off CPU Chips and Beyond}}, 
  year={2020},
  volume={},
  number={},
  pages={749-761},
  doi={10.1109/ISCA45697.2020.00067}}

@article{block_cipher_review,
  author    = {G. Hatzivasilis and K. Fysarakis and I. Papaefstathiou and C. Manifavas},
  title     = {A Review of Lightweight Block Ciphers},
  journal   = {Journal of Cryptographic Engineering (J. Cryptogr. Eng.)},
  volume    = {8},
  number    = {2},
  pages     = {141-184},
  year      = {2018},
  doi       = {10.1007/s13389-017-0160-y}
}

@ARTICLE{peri_area,
  author={Kim, Seung Soo and Yong, Soo Kyeom and Kim, Whayoung and Kang, Sukin and Park, Hyeon Woo and Yoon,Kyung Jean and Sheen, Dong Sun and Lee, Seho and Hwang, Cheol Seong},
  journal={Advanced Materials}, 
  title={{Review of Semiconductor Flash Memory Devices for Material and Process Issue}}, 
  year={2022},
  volume  = {35},
  number  = {2200659},
  pages={},
  doi={10.1002/adma.202200659}}

@ARTICLE{overview_nvm_tech,
  author={Zhao, Chun and Zhao, Ce Zhou and Taylor, Stephen and Chalker, Paul R.},
  journal={Nanoscale Research Letters (NRN)}, 
  title={{Review on Non-Volatile Memory with High-k Dielectrics: Flash for Generation Beyond 32 nm}}, 
  year={2014},
  volume  = {7},
  number  = {7},
  pages={5117–5145},
  doi={10.3390/ma7075117}}

@misc{adaptive_privacy_ssd,
      title={Adaptive Privacy-Preserving {SSD}}, 
      author={Na Young Ahn and Dong Hoon Lee},
      year={2025},
      eprint={2506.02030},
      archivePrefix={arXiv},
      primaryClass={cs.CR},
      url={https://arxiv.org/abs/2506.02030}, 
}

@InProceedings{iceclave,
    author    = {Kang, Luyi and Xue, Yuqi and Jia, Weiwei and Wang, Xiaohao and Kim, Jongryool and Youn, Changhwan and Kang, Myeong Joon and Lim, Hyung Jin and Jacob, Bruce and Huang, Jian},
    title     = {IceClave: A Trusted Execution Environment for In-Storage Computing},
    booktitle = {IEEE/ACM International Symposium on Microarchitecture (MICRO)},
    month     = {},
    year      = {2021},
    pages     = {199-211}
}
%%%%%%%%%%%%%%%%%%%%%%%%%%%%%%%%%%%%

\end{document}